\documentclass[aps,twocolumn,preprintnumbers,amsmath,amssymb,10pt,a4paper,nofootinbib,prx,showkeys
]{revtex4-2}

\usepackage{graphicx}% Include figure files
\usepackage{dcolumn}% Align table columns on decimal point
\usepackage{bm}% bold math
\usepackage{hyperref}% add hypertext capabilities
\usepackage[mathlines]{lineno}% Enable numbering of text and display math
\usepackage{filecontents}
\makeatletter\@input{workaround.tex}\makeatother

\usepackage{braket}
\usepackage{xcolor}
\usepackage{bbm}
\usepackage{xfrac}

\newcommand{\rd}{\mathrm{d}}

\newcommand{\Ocal}{\mathcal{O}}

\usepackage{xr}
\makeatletter
\newcommand*{\addFileDependency}[1]{% argument=file name and extension
  \typeout{(#1)}% latexmk will find this if $recorder=0 (however, in that case, it will ignore #1 if it is a .aux or .pdf file etc and it exists! if it doesn't exist, it will appear in the list of dependents regardless)
  \@addtofilelist{#1}% if you want it to appear in \listfiles, not really necessary and latexmk doesn't use this
  \IfFileExists{#1}{}{\typeout{No file #1.}}% latexmk will find this message if #1 doesn't exist (yet)
}
\makeatother

\newcommand*{\myexternaldocument}[2]{%
    \externaldocument[#2-]{#1}%
    \addFileDependency{#1.tex}%
    \addFileDependency{#1.aux}%
}
\myexternaldocument{supmat_standalone}{SI}

\begin{document}

\title{A symmetry-based approach to species-rich ecological communities}

\author{Juan Giral Mart\'inez}
 \affiliation{Institut de Biologie de l'École Normale Supérieure, Département de Biologie, École Normale Supérieure, PSL Research University, Paris, France}

\begin{abstract}
Disordered systems theory provides powerful tools to analyze the generic behaviors of high-dimensional systems, such as species-rich ecological communities or neural networks. By assuming randomness in their interactions, universality ensures that many microscopic details are irrelevant to system-wide dynamics; but the choice of a random ensemble still limits the generality of results. We show here, in the context of ecological dynamics, that these analytical tools do not require a specific choice of ensemble, and that solutions can be found based only on a fundamental rotational symmetry in the interactions, encoding the idea that traits can be recombined into new species without altering global features. Dynamical outcomes then depend on the spectrum of the interaction matrix as a free parameter, allowing us to bridge between results found in different models of interactions, and extend beyond them to previously unidentified behaviors. The distinctive feature of ecological models is the possibility of species extinctions, which leads to an increased universality of dynamics as the fraction of extinct species increases. We expect that these findings can inform new developments in theoretical ecology as well as for other families of complex systems. 
\end{abstract}


\maketitle

\section{Introduction}
\label{sec:intro}
Ecological communities are archetypical complex systems, with a vast number of degrees of freedom and a variety of processes and scales. Central to ecology, the question of how interactions between species impact their abundances and thus lead to a variety of patterns and dynamics is made difficult by the amount of parameters required to model such a system.\\

Building upon tools from spin-glass theory, this problem has recently seen great theoretical progress, with a focus on reducing the number of parameters to a tractable one, that allows for analytical predictions and for a comprehensive study of the impact of each parameter. The main idea, which was pioneered by May in ecology, is to substitute as much as possible of the complexity of the system by randomness\cite{Gardner1970,May1972}. This approach, rooted in the success of so-called 'disordered models' in theoretical physics, relies on drawing species interactions from some probabilistic model with pre-defined statistics. One then finds that, despite each sample being different, collective properties (i.e. those defined at the level of the whole community) are \textit{generic}, that is, dependent only on the few statistical parameters that characterize the random model. Hence one can predict important features such as Species Abundance Distributions (SAD) \cite{Bunin2017}, ecosystem stability \cite{Galla2018} and the nature of dynamical regimes, without the need to resolve interactions at the species level \cite{Altieri2021}.\\

Whenever there is little ecological information, disordered models may thus be seen as null models that allow to test what can generically be expected from a community \cite{Barbier2018}. In other words, they are useful in the study of \textit{unstructured communities}, those in which no species plays a distinctive role and no particular process can be singled out as having special importance in shaping the dynamics.\\

Most theoretical work so far has revolved around the so-called \textit{random generalized Lotka-Volterra model} (gLV), where interactions are sampled independently from each other \cite{Bunin2017}. Despite recent qualitative agreement with some of the predictions of these models \cite{Gore2022}, many features of ecological communities remain elusive to ‘disordered ecology’\cite{Grilli2020}. Whether these mismatches are the signature of non-random structure in the interactions, and what elements should be added to the models to make them more realistic, are important research questions that have fostered developments in several directions \cite{Galla2023a,Galla2023b,Bunin2022,poley2023,azaele}.\\

Yet, because current random models make stringent assumptions about the nature of the interactions, it can be hard to evaluate which of their conclusions are generic and which are model-specific, a key step to determine when structure is necessary to capture the features of real communities. The goal of this work is to propose a random model for unstructured communities that makes fewer assumptions and is therefore more general.\\

Inspired by physics, our approach consists in defining unstructured communities as those satisfying a natural symmetry. Intuitively, this symmetry must correspond to the idea that, in a community where no species stands out as performing any particular function, the relationship between species and interaction drivers can be randomized without altering the statistical properties of the model. For instance, if interactions are mediated by trait-matching (whereby species having similar traits compete more strongly with one another), we treat the traits of individual species as unimportant as long as correlations amongst traits are fixed (e.g. species that have one trait tend not to have another trait). By imposing this symmetry, the outcomes of the model become independent of the fate of any particular species, and are rather determined by which functions are being performed in the community\cite{Doolittle2017,Louca2018}.\\

Surprisingly, this simple requirement is enough to simplify the dynamics up to a point where stationary regimes and stability conditions can be explicitly described. We show that our symmetry can accomodate a breadth of scenarios that is absent from current models without compromising the ability to make analytical calculations.  Our results reveal that different scenarios lead to notable differences in observables such as Species Abundance Distributions (SAD) and responses to perturbations. All in all, this formalism provides a unified framework to understand what facets of species interactions shape the richness and stability of the community.\\

The paper is structured as follows: in Section \ref{sec:model} we recall the Lotka-Volterra equations and justify, in ecological terms, the symmetry underpinning the random matrix ensemble from which interactions will be sampled. In Section \ref{sec:dmft} we give an overview of the method to simplify the dynamics and analyze the properties of the Unique Fixed Point (UFP) solution. In Section \ref{sec:assembly} we investigate how the extant community is related to the initial pool of species as well as the possibility of dynamical regimes beyond the UFP solution. We conclude in Section \ref{sec:disc}.

\section{Model definition}
\label{sec:model}
\subsection{The Lotka-Volterra equations}
\label{sec:glv}
We consider a set of $S$ interacting species, labeled by $i=1,\dots,S$ and model the temporal dynamics of their respective abundances, which we denote $x_i(t)$, according to the generalized Lotka-Volterra equations (gLV),
\begin{equation}
    \frac{\rd x_i}{\rd t} = r_i x_i \left[1-u_i x_i + \sum_j A_{ij} x_j\right],
    \label{eq:1_1_glv}
\end{equation}
where $r_i$ and $u_i$ are respectively the bare growth rate and the self-regulation strength of species $i$. These parameters are intrinsic parameters of the species: $r$ depends on its adaptation to the environment, whereas $u$ encodes the competition between its members. To reduce the number of parameters, and consistently with our goal of modelling unstructured ecosystems, we will suppose that these quantities are the same for all species. In that case, $r$ can be absorbed by rescaling the time variable, so we will set $r = 1$. On the other hand, we keep $u$ as a free parameter and will use it as a control parameter later on. All the following analytical work can be done without these simplifications.\\

Apart from its interaction with the environment, the net growth rate of the species depends on its interaction with the rest of the community, as represented by the matrix of interaction coefficients $A_{ij}$. This matrix encodes, in an abstract form, the information about all ecological processes that are relevant to understand the dynamics. As such, specifying the special characteristics of a community is tantamount to defining an interaction matrix. The interaction matrix alone accounts for $S^2$ parameters. For $S\gg 1$, this is a prohibitively large number. Reducing the number of parameters is critical to recover a useful and tractable model.\\

The generalized Lotka-Volterra equations are the simplest non-linear model capturing two defining aspects of community ecology: exponential growth in the absence of interactions (due to reproduction) and stability of extinctions (that is, the impossibility for a species gone extinct to re-invade a community without migration from outside). Although seemingly simple, these two characteristics are in fact crucial to the understanding of many facets of the gLV dynamics \cite{Arnoulx,BuninEntropy}.

\subsection{Our definition of structure}
\label{sec:struct}
In order to define a matrix ensemble that represents unstructured communities, a fundamental question must first be tackled: how is structure reflected in the properties of the interaction matrix? This is a vast question, and many independent directions have been explored that lead to different results. Examples include hierarchical interactions \cite{Galla2023a}, food webs \cite{Allesina2015} and sparse networks \cite{Bunin2022,Bunin2023}.
Here, our starting point will be that structure manifests itself in the form of atypical directions in the interaction matrix. For instance, \cite{poley2023} has recently studied communities where interactions are constructed as the superposition of a random matrix and a non-random anisotropic matrix. The resulting non-random directions encode linear combinations of species abundances to which the dynamical system is particularly sensitive. Ecologically, these can be pictured, for instance, as ecosystemic functions (such as metabolite production or consumption) that drive the interactions and influence the dynamics.\\

This leads us to a definition of unstructured interactions as those that don't possess such privileged directions. Mathematically speaking, we require the matrix ensemble to be invariant under orthogonal transformations,
\begin{equation}
    \forall U\in \Ocal_S,  \rd \mathbb{P}\left(UAU^T\right) = \rd \mathbb{P}(A),
    \label{eq:1_2_invariance}
\end{equation}
where $\mathbb{P}$ is the measure associated to the matrix ensemble and $A$ is any matrix belonging to its support. Such models can be obtained by fixing an interaction matrix $A_0$ and rotating it by a random orthogonal matrix,
\begin{equation}
    A = U A_0 U^T, ~ U\sim \mathrm{Haar}\left(\Ocal_S\right),
    \label{eq:1_2_inv_model}
\end{equation}
the Haar measure being the uniform probability measure over the orthogonal group. Some ecological insight can be gained by inspecting what happens when $A_0$ is transformed into $A$ in such a way. Suppose that $A_0$ is decomposed into a sum of rank-one components that represent the different processes shaping interactions,
\begin{equation}
    A_0 = \sum_\lambda u_\lambda v_\lambda^T,
    \label{eq:1_2_funct_decomp}
\end{equation}
which naturally arise in consumption/cross-feeding models \cite{MacArthur1955,Advani2018} (in which case $u_\lambda = v_\lambda$ is the preference vector associated to resource $\lambda$, i.e. the rate at which species feeds on/produces that resource), trait-based models \cite{Galla2023b} and function-based models (in which case $u_\lambda$ and $v_\lambda$ respectively encode the species's response to and effect on function $\lambda$). Conjugating $A_0$ by the orthogonal matrix $U$ amounts to randomly rotating each of the vectors $u_\lambda$, $v_\lambda$, thereby randomizing the relationship between species and resources (resp. traits or functions). This randomization is consistent with the idea of no species playing a well-defined role in an unstructured community. Importantly, because rotations preserve scalar products, quantities such as $u_\lambda \cdot u_\nu$ are preserved, meaning that the overlaps between the important directions of the matrix are preserved, despite the directions themselves being randomized.\\

Mathematically, the effect of conjugating by a matrix $U$ can be pictured as a sort of mixing, whereby a new pool of species is defined whose interactions are a blend of the interactions of the previous pool. Our model is defined so that any such mixture is equally likely, as long as it preserves correlations between interaction drivers. In other words, it is \textit{statistically invariant} under mixing. At the same time, due to the large number of species, typical realizations of the system will correspond to typical realizations of $\mathrm{Haar}\left(\mathcal{O}_S\right)$. This rules out eigenvector anisotropy (which may correspond, for instance, to a non-zero mean interaction or a division of the community into several groups), eigenvector localization (which rules out the presence of sub-communities made of strongly interacting species, as in some sparse models) and correlations between eigenvalues and eigenvectors.\\

Once randomization is done, the only remaining information about the original interaction matrix are its eigenvalues, which are common to the matrices of the ensemble thus defined. In particular, the spectrum encodes how the interactions were originally constructed. Importantly, the second moment of the spectral distribution determines the variance of the interaction coefficients, which equals $\langle \lambda^2 \rangle/S$.\\

In the simplest case, the spectrum is semi-circular, and the ensemble considered is the Gaussian Orthogonal Ensemble (GOE), where the $\left(A_{ij}\right)_{i < j}$ are jointly distributed as independent Gaussians, and $k$-point statistics of the form $\langle A_{i_1j_1}\dots A_{i_k j_k} \rangle$ vanish at any order $k$. For other spectral distributions, we show in Appendix \ref{sec:app_coefs} that such statistics still vanish asymptotically for $k \ll S$, but don't when the number of terms is comparable to the size of the system. Thus coefficients appear to be independently Gaussian when taken in small sets but not when considering the whole matrix. This diffuse pattern of high-order inter-dependencies thus has an impact when considering quantities that depend on an extensive number of interaction coefficients, such as the terms $\sum_j A_{ij} x_j$ in \eqref{eq:1_1_glv}. Thus they have an impact on the dynamics, as we show next.

\section{Dynamical mean-field theory (DMFT)}
\label{sec:dmft}
\subsection{General setting}
\label{sec:dmft_intro}
In this section, we reduce the complex high-dimensional dynamics of \eqref{eq:1_1_glv} to a one-dimensional, albeit stochastic equation, assuming the prescription \eqref{eq:1_2_inv_model} for the interaction matrix. For readability purposes, we restrict our analysis to symmetric matrices, in which case formulas are more compact and the number of parameters is reduced. We provide a way to solve the dynamics for non-symmetric matrices, as well as other straightforward extensions of the model, in Supplementary Information (see SI Sec. \ref{SI-sec:sparse} \& \ref{SI-sec:asym}).\\

Under the assumption that $A$ is symmetric, $A_0$ can be diagonalized in an orthonormal basis with real eigenvalues, so that without loss of generality $A_0$ can be taken to be diagonal. The matrix ensemble is then characterized by the distribution of its eigenvalues, whose \textit{probability density function} (pdf) we denote $\rho$, so that the unstructured model can be divided in as many classes as there are pdfs. We further suppose that the support of $\rho$ is bounded and doesn't scale with $S$, thus implying that our gLV model is in the so-called \textit{weak interaction regime}, where interactions are weak when taken individually but strong at the collective level\cite{Giulia22}. We emphasize that, at odds with traditional uses of orthogonal ensembles in Random Matrix Theory (RMT), the spectral distribution need not be a random object, but rather a parameter that encodes the way interactions are driven.\\

Since $A_{ij}$ are random variables, the dynamical trajectories will be random as well: our goal is to characterize the law of the random processes $x_i(t)$. To this end, we use the well-known path integral formalism \cite{Hertz2017}, that starts by rewriting the probability of a given outcome $\Vec{x}(t)$ as an integral over all possible realizations of the random matrix,
\begin{equation}
    \rd p\left(\Vec{x}(0)\right) \int \rd \mu (U) \prod_i\delta \left(\frac{\dot{x_i}}{x_i} - 1 + u x_i - \sum_j A_{ij} x_j\right),
    \label{eq:3_1_path_integral}
\end{equation}
where $\mu$ stands for the Haar measure over the orthogonal group, $p$ is the distribution of the initial conditions (which might be deterministic or random) and $\delta$ is a functional Dirac delta that enforces the trajectory to undergo the prescribed dynamics. The goal is then to simplify this expression up to a point where the probability distribution $\rd \mathbb{P}\left[x_i(t)\right]$ can be recognized explicitly in a simple form. The strategy to do so is standard in disordered systems\cite{Opper1989,Galla2018}, hence we mainly focus on the results, and relegate details to SI (see SI Sec. \ref{SI-sec:dmft}). The key step, performing the integral over $U$, relies on the following large deviation principle for the so-called HCIZ integral (see \cite{HCIZ,Marinari1994, Zuber80, PottersBouchaud} as well as SI Sec. \ref{SI-sec:free}),
\begin{equation}
    \int \rd \mu(U) \exp\left[\frac{S}{2} \mathrm{Tr} \left(U A_0 U^T B\right)\right] \sim \exp\left[\frac{S}{2} \mathrm{Tr} G_\rho \left(B\right)\right],
    \label{eq:3_1_HCIZ}
\end{equation}
where $A_0$ is the aforementioned diagonal matrix and $B$ can be any low-rank symmetric matrix of bounded spectrum. This introduces the function $G_\rho$, usually called \textit{free cumulant generating function}, whose detailed definition can be found in Appendix \ref{sec:app_G}. In short, $G_\rho$ characterizes the shape of $\rho$. It will play a key role in determining the dynamics.\\

In practice, the averaging step is where the symmetry assumption is introduced in the model. This causes the path integral to no longer depend on all interaction coefficients but only on the spectrum $\rho$ and the two macroscopic quantities,
\begin{equation}
    \begin{split}
        C(t,s) &= \braket{x(t) | x(s)} = \frac{1}{S} \sum_i x_i(t)x_i(s)\\
        K(t,s) &= \left\langle \text{\Large$\sfrac{\delta}{\delta \xi(s)}$}\Big|x(t)\right\rangle= \frac{1}{S} \sum_i \frac{\delta x_i (t) }{\delta \xi_i (s)}.
    \end{split}
    \label{eq:3_1_order_parameters}
\end{equation}
Both of them are descriptors of the dynamics. Indeed, $C$ is the auto-correlation of the abundance time-series, averaged over all species. $K$ is the average response of abundances at a given time $t$ w.r.t. an environmental perturbation at a previous time $s$ (see Eq. \eqref{eq:3_1_effective_sde} for the definition of $\xi$). In \eqref{eq:3_1_order_parameters} we have purposefully written them as scalar products in the space of abundances to reveal the fundamental role of orthogonal invariance: once the symmetry is averaged out, the path integral (and hence the dynamics) may only depend on quantities that are invariant under rotations. In other words, it may only depend on scalar products in the space of species abundances. By definition, these are collective quantities, defined as averages over all species. The fact that the path integral may only depend on them is what generates a regime of \textit{collective dynamics}.\\

Upon subsequent simplifications, the path integral can finally be identified as generating the set of Stochastic Differential Equations (SDE),
\begin{equation}
    \frac{\rd x_i}{\rd t} = x_i \left[1 - u x_i + \int_0^t H(t,s) x_i(s) + \xi_i(t)\right].
    \label{eq:3_1_effective_sde}
\end{equation}
Each species $i$ effectively follows the dynamics prescribed by this process independently of the other species. In particular, the effect of inter-species interactions has been replaced by two terms. First, a memory term driven by a \textit{memory kernel} $H(t,s)$: it introduces a time-delayed self-regulation of the species on itself. Second, a Gaussian noise $\xi_i(t)$ characterized by a \textit{noise correlator} $C_\xi(t,s)$. The fact that $\xi_i$ has Gaussian statistics can be traced back to the fact that, as was noted in the previous paragraph, scalar products are the only invariant quantity under orthogonal symmetry. In other words, the type of noise that is obtained in the effective equation is a direct consequence of the symmetries of the model.\\

Both $H$ and $C_\xi$ are deterministic functions that can be obtained from $C$ and $K$. In fact, $C$ and $K$ are also deterministic functions, since \eqref{eq:3_1_effective_sde} defines them as averages over independent species. This implies a self-consistency relation between the effective drivers of single-species dynamics ($H$ and $C_\xi$) and the empirical many-species statistics ($C$ and $K$).\\

In principle, these self-consistency relations (which we relegate to SI Sec. \ref{SI-sec:dmft} for readability) would allow to solve for the four quantities. In practice, they include non-linear functional operators and are too intricate to be solved. Even a numerical solution would in fact prove far more costly than the already intensive solution proposed for GOE matrices in \cite{Roy2019}. Hence, in the following we focus on stationary regimes, where most of this complexity fades away.

\subsection{Stationary regimes}
\label{sec:dmft_stationary}
We define stationary regimes as those where one-point averages are independent of time and two-point averages depend only on time differences, i.e. $C(t,s) = C(t-s)$ etc. Previous works have leveraged the assumption of stationarity, even in conditions where the community doesn't reach a fixed point \cite{RoyThesis}. Fortunately, time-translation invariance is enough to get rid of the most complex part of the general analysis, namely the fact that the self-consistent closure equations come in the form of functional operators. Indeed one can then rewrite the closure equations as scalar relationships between the Fourier modes of the different quantities. Let hatted letters denote the Fourier transforms of the corresponding quantities. Then we find (see SI Sec. \ref{SI-sec:stat})
\begin{equation}
    \begin{split}
        \hat{H}(\omega) &= G_\rho'\left(\hat{K}(\omega)\right) \\
        \hat{C_\xi}(\omega) &= \hat{C}(\omega) \frac{\mathrm{Im} \left[G_\rho'\left(\hat{K}(\omega)\right)\right]}{\mathrm{Im} \left[\hat{K}(\omega)\right]}.
    \end{split}
    \label{eq:3_2_Fourier_closure}
\end{equation}
The first equation gives a mapping between the response function and the memory kernel. The second relates the fluctuations in population abundances to the DMFT noise fluctuations. In particular, the value of $\hat{K}(\omega)$ appears as a modulator to the relationship between abundance and noise fluctuations, suppressing or enhancing frequencies (but not de-phasing) depending on $G_\rho'$. This is in contrast to the GOE case studied in \cite{Bunin2017}, where the two are directly proportional. We interpret this as a signature of memory in the system due to the diffuse correlations present in the interaction matrix.

\subsection{Fixed points}
\label{sec:dmft_fp}
We now specialize the general solution even more to the case where the dynamics reach a fixed point. Compared to time-translation invariant regimes, this is obtained by further requiring that correlation functions are also time-independent. Loss of time dependence in the correlators ensures that all random and non-random processes converge to a fixed point. In particular, we define
\begin{equation}
    \begin{split}
        x(t) &\rightarrow x\\
        \xi(t) &\rightarrow \xi\\
        C(t,t) &\rightarrow q\\
        C_\xi(t,t) &\rightarrow z\\
        \int_0^t \rd s K(t,s) &\rightarrow \chi \\
        \int_0^t \rd s H(t,s) &\rightarrow \eta .
    \end{split}
    \label{eq_3_3_fp_variables}
\end{equation}
The first two are random quantities. In particular, $\xi$ is a centered Gaussian variable with variance $z$. On the other hand, $q$, $z$, $\chi$ and $\eta$ are defined as limits of deterministic processes and are therefore deterministic. More precisely, $q$ is the mean-square abundance at equilibrium, related to the variance of the abundance distribution; $\chi$ is the integrated response function, that is the response of the equilibrium abundance of a species to a permanent change in its carrying capacity; and $\eta$ is the integrated memory kernel, i.e. a modifier to the species self-regulation due to feedback through its interactions with other species. By setting the derivative to zero in \eqref{eq:3_1_effective_sde}, one obtains the relationship between $x$ and $\xi$,
\begin{equation}
    x = \mathrm{max}\left(0,\frac{1 + \xi}{u-\eta}\right).
    \label{eq:3_3_fp_sad}
\end{equation}
The species abundance distribution is therefore a truncated Gaussian, extending known results for Wigner and Wishart matrices \cite{Bunin2017,Galla2023b}.\\

The fixed point equations relating the deterministic quantities can then be obtained by taking the limit $\omega \rightarrow 0$ in the Fourier closure \eqref{eq:3_2_Fourier_closure}, rewriting $\eta = \hat{H}(0)$, $q = \hat{C}(0)$, $\chi = \hat{K}(0)$ and $z = \hat{C}_\xi(0)$, and making use of \eqref{eq:3_3_fp_sad} to explicitly write down the resulting averages over $x$. Details of the derivation can be found in SI Sec. \ref{SI-sec:fp}. They read
\begin{equation}
    \begin{split}
    \eta &= G'_\rho \left(\chi\right) \\
    z &= q G''_\rho \left(\chi\right) \\
    q &= \langle x^2 \rangle = \frac{z}{(u-\eta)^2} \omega_2\\
    \chi &= \frac{\omega_0}{u-\eta} .
    \end{split}
    \label{eq:3_3_fp_closure}
\end{equation}
In writing them we have introduced the auxiliary variables $\omega_k = \int_{-1/\sqrt{z}}^\infty \mathcal{D}y \left(y + 1/\sqrt{z}\right)^k$, where the integration is with respect to the standard Gaussian measure. In particular, $\omega_0$ is the fraction of surviving species at equilibrium. Eqs. \eqref{eq:3_3_fp_closure} can also be obtained using the replica trick \cite{SGB_book,Altieri2021}, although in that case one doesn't obtain the dynamics. Inspecting \eqref{eq:3_3_fp_closure} immediately reveals that there are only two independent variables: for instance one may only keep the dynamical descriptors $q$ and $\chi$ or rather take the DMFT quantites $z$ and $\eta$. We keep the four equations to make them more readable.\\

Known results can be obtained by specializing $\rho$ to previously studied distributions. For instance, if $\rho$ is the semi-circle distribution with variance $\sigma^2$ then $G_\rho(x) = \sigma^2 x^2 / 2$ and one obtains the well-known equations for the GOE case \cite{Altieri2021,Bunin2017,Galla2018}. For other spectral distributions, $G_\rho$ contains higher-order cumulants that influence equilibrium properties. In Sec. \ref{sec:struct} we identified such higher-order terms as a signature of inter-dependencies between interactions at the level of the whole matrix. Here, we see how they impact the dynamics at the collective level, through their influence on aggregated quantities such as $z$ and $\eta$.\\

In the GOE case, the appearance of a Gaussian random variable in \eqref{eq:3_3_fp_sad} can be understood from a cavity-like argument \cite{Bunin2017}: since interactions are (up to symmetry) independent from each other, and species are uncorrelated, the interaction term in \eqref{eq:1_1_glv} converges to a Gaussian in the limit of a large number of species. Moreover, the Central Limit Theorem (CLT) applies and gives the relation $z=\sigma^2 q$. For other spectra, interactions are no longer independent, and hence the CLT cannot be applied. In spite of this, the interaction sum still falls in the basin of attraction of a Gaussian distribution, but the relation between $z$ and $q$ in Eq. \eqref{eq:3_3_fp_closure} is modified to take correlations into account. Hence our ensemble operates in an intermediate regime between uncorrelated and strongly correlated interactions.\\

In the following, we illustrate the influence of $G_\rho$ by taking three reference classes of spectra. The first class is made of shifted and scaled Marcenko-Pastur distributions, which are associated to the Hebbian matrices studied in \cite{Galla2023b}. These matrices appear when interaction are driven by trait-matching or resource consumption. The second is made of uniform distributions over bounded intervals. The last is made of binary distributions, where the eigenvalues are evenly split between two values $\pm\lambda$. The last two examples are taken purely as theoretical tools for illustration purposes, without well-established theoretical scenarios in mind. The functions $G_\rho$ associated to each of these cases are given in Appendix \ref{sec:app_G}.\\

Fig. \ref{fig:comparison_sad} shows the Species Abundance Distribution (SAD) for these three classes of spectra. Thus . Importantly, the spectra were tailored so as to all have the same variance. Therefore the variance in the interaction coefficients is the same for all three classes. The differences in the results are therefore due to the higher-order cumulants encoded in $G_\rho$ or, in other terms, to the different shapes of each spectrum. By considering only the variance of the interactions and neglecting the shape of the spectra, a random model based on the GOE ensemble would miss these differences.\\

As a last remark, we note that our results build a bridge between the dynamical point of view of DMFT and the point of view of Random Matrix Theory, where one is often interested in computing the spectrum given a random matrix model. By leveraging our formalism, one can directly translate RMT outputs into Lotka-Volterra dynamics.
\begin{figure}
    \centering
    \includegraphics[width=\linewidth]{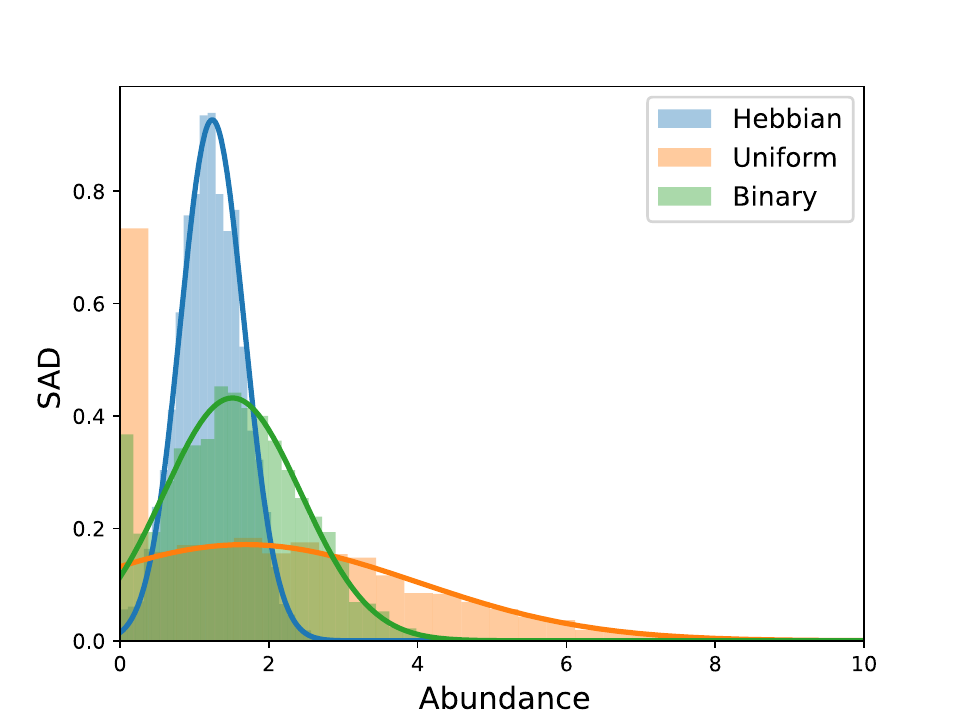}
    \caption{\textbf{Influence of the spectral distribution on equilibrium abundances: } The Species Abundance Distribution (SAD) is plotted for three different spectra (bars are simulations, lines are predictions). \textit{Blue:} A shifted and scaled Marcenko-Pastur distribution, corresponding to Hebbian interactions. \textit{Orange:} A uniform distribution. \textit{Green:} A binary distribution, with only two unique values. The spectral distributions all have a variance of $0.4$ and the self-regulation is set to $u=1$, but the ensuing SADs are notably different.}
    \label{fig:comparison_sad}
\end{figure}
\section{Dynamical regimes and reduced interactions}
\label{sec:assembly}
In this section, we investigate how the dynamical process selects the interactions, and give conditions for the stability of the fixed point solution found in \ref{sec:dmft_fp}. Finally, we discuss the possible dynamical regimes beyond the fixed point solution and their relationship to orthogonal symmetry.
\subsection{Spectrum of the reduced community}
\label{sec:assembly_spectrum}
As the system reaches a fixed point, some species remain and some go extinct. We let $S^\star$ be the number of surviving species and $A^\star$ the $S^\star \times S^\star$ matrix obtained by restricting $A$ to surviving species. We call this the \textit{reduced interaction matrix}. Even though species are, to leading order in $S$, independent from each other, subtle correlations exist that tend to favor species experiencing less competition. As such, $A^\star$ is not just a generic sub-matrix of the full interaction matrix. In particular, it has been shown that reduced matrices tend to exhibit in-row and in-column correlations \cite{Bunin2016}. On the other hand, \eqref{eq:3_3_fp_closure} clearly shows that the properties of the fixed point, which are related to the properties of the reduced matrix, depend solely on $\rho$, thereby hinting at a simple relationship between the full spectrum and the assembled spectrum. In fact, one can check that \eqref{eq:3_3_fp_closure} is invariant under the following transformation
\begin{equation}
    \begin{split}
        G_\rho(x) &\longrightarrow \omega_0^{-1} G_\rho(\omega_0 x) \\
        \omega_0 &\longrightarrow 1.
    \end{split}
    \label{eq:4_1_G_transformation}
\end{equation}
Heuristically, we interpret this as the fact that the equilibrium can be seen both from the point of view of an interaction matrix with spectrum $\rho$ leading to a community with a surviving fraction $\omega_0$ or from the point of view of a community where all species survive but with spectrum $\nu$ such that $G_\nu(x) = \omega_0^{-1} G_\rho(\omega_0 x)$.  This is of course a heuristic interpretation, rather than a rigorous analytical argument, since $\omega_0$ is an outcome of the dynamical process, not a free parameter. Yet, numerical results under a variety of spectral distributions (semicircular, Marcenko-Pastur, uniform, Beta, and power distributions) confirm that the spectrum of $A^\star$ is indeed distributed according to $\nu$ so defined.\\
\begin{figure*}[t]
    \centering
    \includegraphics[width=16cm]{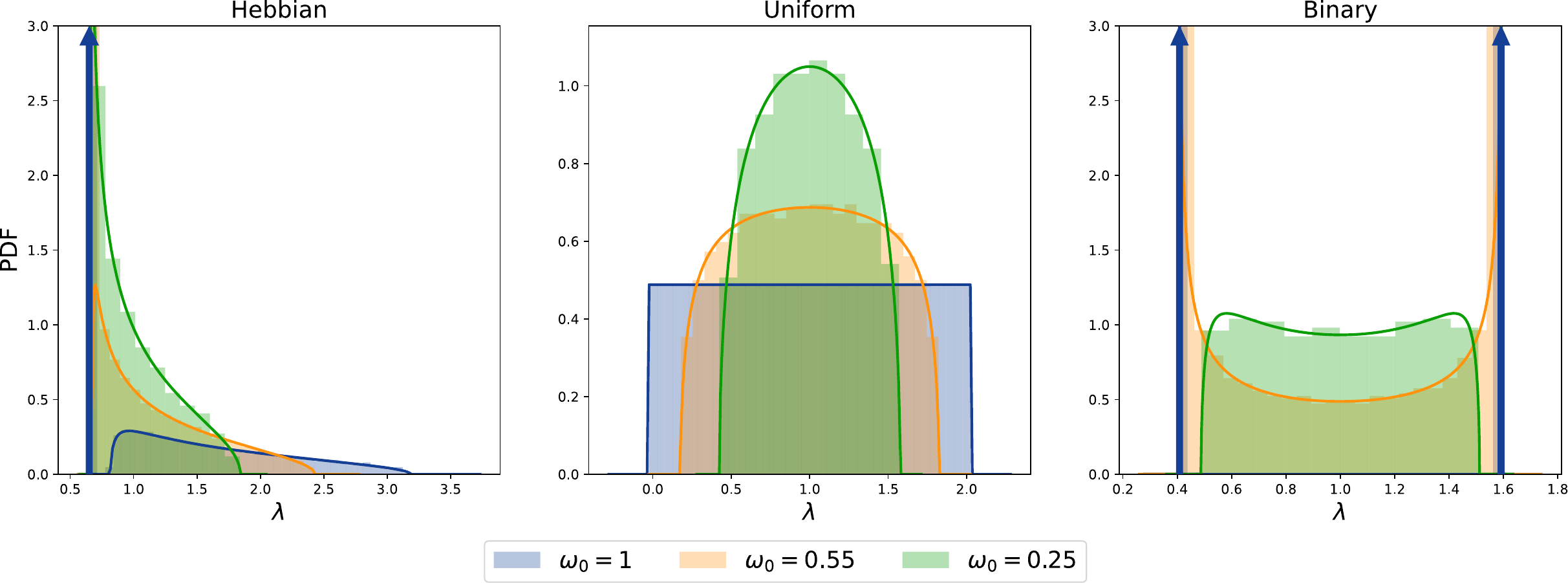}
    \caption{\textbf{Influence of $S^\star / S$ on the assembled spectrum.} The spectra of typical submatrices are shown for three classes of orthogonal models (Hebbian, Uniform and Binary, same as in Fig. \ref{fig:comparison_sad}) and different survival fractions. Blue curves correspond to the spectrum of the full matrix. Histograms correspond to numerical simulations and solid curves are predictions using Eq. \eqref{eq:4_1_G_transformation}. Both the Hebbian and the binary case have Dirac masses for $S^\star / S = 1$, which are signalled by arrows. As $S^\star / S$ is decreased, spectral distributions drift away from their initial shape and towards a semi-circular distribution. In particular, the edges of the spectrum are softened to match the continuous edges of the semi-circular distribution.}
    \label{fig:spectra}
\end{figure*}

Interestingly, we show (see SI Sec. \ref{SI-sec:red-mat}) that $G_\nu(x) = \omega_0^{-1} G_\rho(\omega_0 x)$ also characterizes the spectral density of a generic submatrix of size $S^\star = \omega_0 S$ of the original matrix. In other words, even though the reduced matrix has intricate correlations and doesn't belong to an orthogonal ensemble anymore, the bulk of its spectrum matches what one would get by randomly sampling a set of $S^\star$ species and looking at their interactions, therefore bypassing the whole dynamical process. This was known for GOE matrices, where the reduced matrix also displays a semi-circular bulk up to a rescaling of the variance to $\sigma^2 \omega_0$.\\

Yet, for general spectral distributions, the transformation induced by \eqref{eq:4_1_G_transformation} is more subtle than a simple rescaling of the variance, and $A^*$ will not necessarily belong to the same class of matrices as $A$. In particular, given that higher-order cumulants are scaled down, communities with a lower survival rate tend to have a spectrum that 'interpolates' between the original spectrum and the semi-circle distribution. This is shown in Fig. \ref{fig:spectra} for our three reference classes of spectra.
\subsection{Linear instability}
\label{sec:assembly_stability}
The results in \ref{sec:dmft_fp} and \ref{sec:assembly_spectrum} have been derived under the assumption that a unique stable fixed point (UFP) exists. This is trivially true when species don't interact (that is, $\rho(x) = \delta(x)$) due to the self-regulation $u$ of the species. As the spectrum grows and the interactions become stronger, the stable fixed point solution remains until an instability threshold is crossed \cite{Bunin2017,Opper1989}. In this section, we characterize the location at which this transition happens and the dynamics beyond the transition.\\

We probe the stability of the fixed point by introducing a small  perturbation to the fixed point abundances \eqref{eq:3_3_fp_sad} and studying the stability of the linearized SDE. This is equivalent to checking the stability of the Jacobian.\\
Let $\{x_i^\infty\}$ be an equilibrium configuration where each species is an independent realization of \eqref{eq:3_3_fp_sad}.  Let $\delta x_i (0)$ be a small non-random perturbation to those equilibrium abundances.
We let $x_i(t) = x_i^\infty + \delta x_i(t)$ be the ensuing dynamics. Because the system is no longer at equilibrium, the DMFT noise will also acquire a fluctuating part, which self-consistently stems from the fluctuations in the abundances. We denote it $\xi_i(t) = \xi_i^\infty + \delta \xi_i(t)$. Assuming that $\delta x_i(t)$ and $\delta \xi_i(t)$ remain small, we linearize the full dynamics \eqref{eq:3_1_effective_sde} to get
\begin{equation}
\begin{split}
    \frac{\rd}{\rd t} \delta x_i &= x_i^\infty \Big[-(u-\eta) \delta x_i(t) + \delta \xi_i(t)\Big]\\ 
    &+ \delta x_i(t) \Big[1 - (u-\eta)x_i^\infty + \xi_i^\infty\Big].
\end{split}
\label{eq:4_2_linearized}
\end{equation}
As per the equilibrium conditions, the second term in the r.h.s is zero for extant species and negative for extinct species. It can therefore never cause an instability. The first term is zero for extinct species. Therefore we may focus on this term and keep track of the surviving community only. In particular, instabilities may never originate in the extinct community.\\

Just as for the full dynamics, $\delta \xi_i(t)$ is a Gaussian process. Because the dynamical equation is now linear, this implies that $\delta x_i(t)$ is also a Gaussian process. In fact, it is convenient to take the Fourier transform of \eqref{eq:4_2_linearized},
\begin{equation}
    \left[i\omega + x_i^\infty\left(u -\eta\right)\right] \hat{\delta x_i}(\omega) = x_i^\infty \hat{\delta \xi_i}(\omega) + \delta x_i(0).
    \label{eq:4_2_laplace}
\end{equation}
Instabilities are signalled by resonances in this expression. In particular, given that $\hat{\delta x_i}(\omega)$ is a Gaussian variable, we may look for divergences of its mean and its variance. After some manipulations (see SI Sec. \ref{SI-sec:instab}) we find that the mean is always stable and that the variance diverges when the following condition is met:
\begin{equation}
    \left(u-\eta\right)^2 - \omega_0 G_\rho''\left(\frac{\omega_0}{u-\eta}\right) = 0.
    \label{eq:4_2_sq_avg_instability}
\end{equation}
Note how this condition is consistent with the spectral transformation \eqref{eq:4_1_G_transformation}. Leveraging this, we show (see SI Sec. \ref{SI-sec:mat-inst}) that the onset of instability happens precisely when
\begin{equation}
    u - \lambda_+\left(A^\star\right) = 0,
    \label{eq:4_2_bulk_instability}
\end{equation}
where $\lambda_+$ is the upper edge of the bulk of $A^*$. That is, the stability of the fixed point is equivalent to the stability of the reduced interaction matrix, similar to what happens in the Wigner case.\\

Interestingly, \eqref{eq:4_2_sq_avg_instability} not only implies a linear instability of the fixed point, but also a divergence of the mean square abundance $q$ (see SI Sec. \ref{SI-sec:is-div}). In other words, all fixed point instabilities lead to a Diverging Abundance (DA) phase, where a finite fraction of species's abundances go to infinity in finite time. The relevance of such dynamical phase is discussed in Sec. \ref{sec:dyn_reg}.\\

Despite always leading to a DA phase, the transitions signalled by \eqref{eq:4_2_sq_avg_instability} can be of two different types, depending on the behavior of $u-\eta$. In the first case, $u-\eta \rightarrow 0$, which is associated to a divergence of the response function $\chi$, as defined in \eqref{eq:3_3_fp_closure}. In the second, $\eta < u$ and $\chi$ remains finite. The second case corresponds to the usual divergence observed in the Gaussian case, as it is indeed the only possible one under the assumption that $\rho$ is semi-circular.\\

Phenomenologically, finite $\chi$ transitions occur when the variance of the DMFT noise becomes too strong to be countered by the self-regulation of the species. At this point, some extensive subset of the community has strong-enough mutualistic interactions to produce a runaway of their abundances to infinity. In other words, the transition is caused by the feedback loop between $q$ and $z$ and causes $q,z \rightarrow \infty$. This always happens with a survival fraction $\omega_0 = 1/2$. On the other hand, infinite $\chi$ transitions are characterized by a vanishing effective self-regulation $u_\mathrm{eff} = u - \eta$. Since species are thus no longer subjected to competition from their own kind, they become able to grow exponentially even at large abundances, hence causing $q\rightarrow\infty$ again. Here, however, $z$ remains finite, i.e. the runaway is not caused by the feedback between $q$ and $z$. When this happens, more than half of the species survive, $\omega_0 > 1/2$.\\

This difference in behavior can be traced back to the shape of the reduced interaction matrix at the onset of instability. Whenever the spectrum of $A^*$ vanishes fast enough near the edge, a finite $\chi$ transition happens. This is the case, for instance, when the spectral density goes as $\nu(\lambda) \sim (\lambda_+ - \lambda)^\delta$ near the edge $\lambda+$. Most continuous spectra (including the semi-circle law, for which $\delta=1/2$) will satisfy this requirement. On the other hand, if $\nu$ doesn't vanish near the edge, an infinite $\chi$ transition occurs. The simplest case scenario here is that of a spectrum having a Dirac mass at the edge. This happens, for instance, if the reduced matrix has a Marcenko-Pastur spectrum with more surviving species than degrees of freedom \cite{Galla2023b}. The impact of the shape of the spectrum on the kind of transition stresses the importance of the transformation given by Eq. \eqref{eq:4_1_G_transformation}. Indeed, as the surviving fraction is decreased from one, the edges of the spectrum are softened, as was noted in Fig. \ref{fig:spectra}. Hence $A^*$ may not have the same edge properties as $A$. In particular, if $A$ already has soft edges then so does $A^*$ and $\chi$ always remains finite at the transition (e.g. Fig. \ref{fig:spectra}, center panel). For $A$ having hard edges, the question is whether these have been softened enough at the onset of instability. This can elucidated by going back to Eq. \eqref{eq:3_3_fp_closure}. It is clear that for a divergent $\chi$ transition to occur $G_\rho'(x) \xrightarrow[x\rightarrow\infty]{} u$ needs to hold, in which case Eq. \eqref{eq:4_2_sq_avg_instability} automatically holds in the limit. This is a condition for the transition to be plausible. For it to occur, we also need to check that a finite $\chi$ transition isn't reached beforehand. Simple manipulations (see SI Sec. \ref{SI-sec:div-type}) lead to the criterion
\begin{equation}
\lim_{x\rightarrow\infty}x^2 G_\rho''(x) > \frac{1}{2}.
\end{equation}
When this is satisfied, the finite $\chi$ transition takes place, otherwise the divergent $\chi$ transition occurs.

\subsection{Global dynamical regimes}
\label{sec:dyn_reg}

For ecological applications, the Diverging Abundances (DA) phase is meaningless. It stems from the fact that, as a phenomenological model, the Lotka-Volterra equations break down when there are sufficiently many positive interactions in the system. Further terms are needed to fix this pathological behavior. For example, a negative coupling to the total abundance (which would correspond to a non-zero mean in the interaction coefficients, biasing them towards competition) can displace (although not fully remove) the divergence of abundances. The interaction matrix then reads
\begin{equation}
    A_{ij} = \frac{\mu}{S} + \Tilde{A}_{ij},
    \label{eq:4_3_regularized_matrix}
\end{equation}
for some $\mu < 0$ and $\Tilde{A}$ satisfying orthogonal invariance. In that case, even though the fixed point becomes unstable, the increased competition as abundances grows prevents the runaway to infinity beyond the linear instability. This region, where the fixed point solution is unstable but the system is non-divergent, has been proven to be a 'Multiple Attractors' (MA) phase by previous works \cite{Altieri2021} in the particular case of a Wigner spectrum. There, the unique fixed point is replaced by an exponentially large number of fixed points\cite{Ros}.\\

\begin{figure*}[t]
    \centering
    \includegraphics[width=16cm]{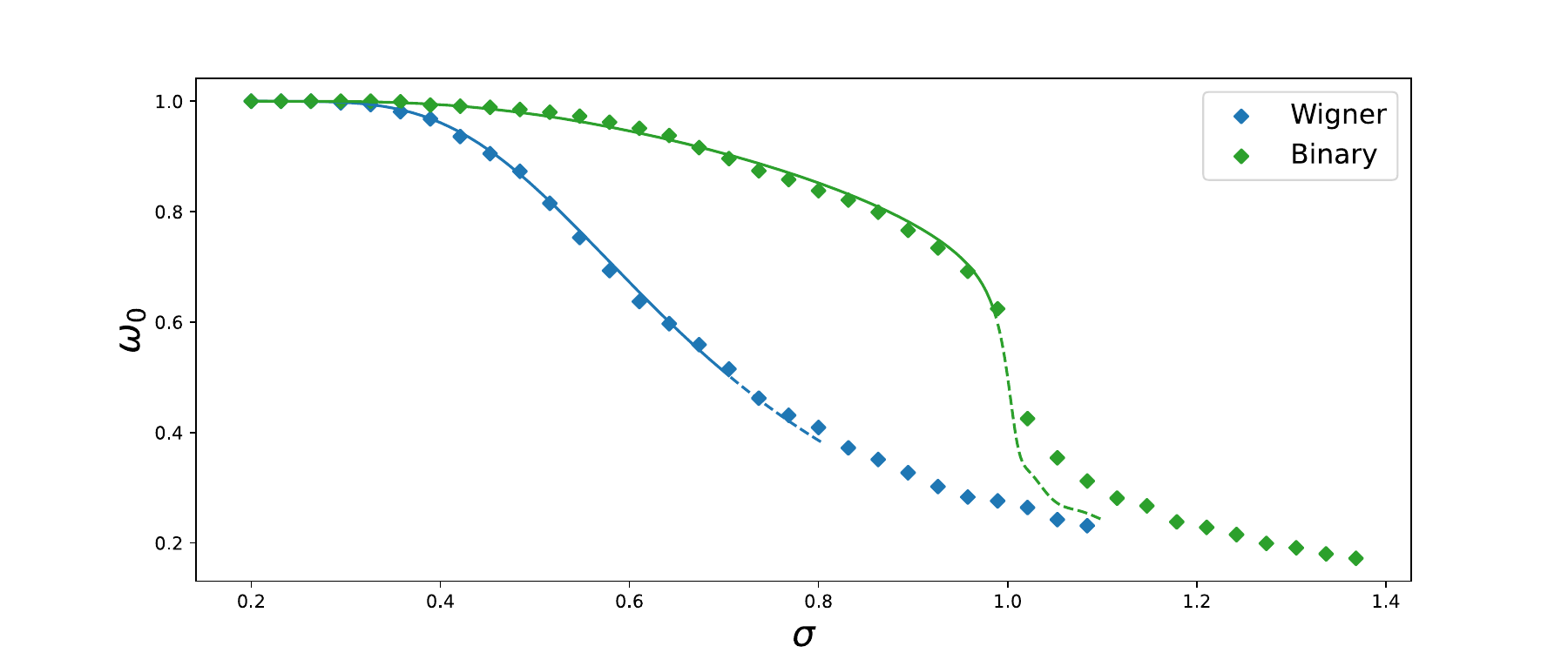}
    \caption{\textbf{Influence of the type of transition on the survival fraction.} Biasing the community towards competition by means of a non-zero average interaction (see \eqref{eq:4_3_regularized_matrix}, here $\mu=-5$) removes the divergence of abundances but keeps the linear instability of the fixed point. This leads to a 'Multiple Attractors' (MA) phase that is absent from the orthogonally-invariant model. Here we show the transition from the Unique Fixed Point (UFP) phase to the MA phase as the second moment $\sigma^2$ of the spectra is increased. Whether $\chi$ is finite or infinite at the instability can be determined from the behavior of the fraction of surviving species $\omega_0$ at and beyond the transition in this regularized model. Finite $\chi$ transitions (as for Wigner spectra, \textit{blue}) are characterized by a continuous decrease of $\omega_0$. Conversely, $\chi\rightarrow \infty$ leads to a discontinuous drop (as for binary spectra, \textit{green}). \textit{Dots} are obtained from numerical simulations with $S=1000$ and \textit{solid lines} are predictions adapting the fixed point equations (see \ref{sec:dmft_fp}) to accomodate $\mu\neq 0$, in the region where the UFP is stable. By extending these predictions to the beginning of the MA phase (\textit{dashed lines}), we see that the now-unstable UFP solution still captures the drop in $\omega_0$.}
    \label{fig:divergences}
\end{figure*}
Orthogonal invariance is the reason why such MA regimes cannot be observed in our model. Indeed, a non-zero mean in the interactions corresponds to an anisotropy along the direction of the total abundance, given by the vector $(1,\dots,1)$. More generally, other low-dimensional anisotropies, such as a division of the community into groups, may lead to interesting phases such as bi-stability or, in the context of non-symmetric interactions, persistent fluctuations \cite{Arnoulx}. In any case, our model shows that these interesting dynamical regimes are a feature of \textit{structured} communities, in the sense that some sort of symmetry breaking is required to get them. Even though we do not pursue this path in the present work, we note that all our calculations can be extended straightforwardly to account for any low-dimensional structure in the interaction matrix. Instead, we restrict ourselves to the simplest symmetry breaking, given by \eqref{eq:4_3_regularized_matrix}, and check how the two transitions identified in \ref{sec:assembly_stability} influence the MA phase beyond the linear instability. Given that the difference between finite and infinite $\chi$ is a local feature (i.e., based on a local stability analysis), we expect the main features identified in the following to remain meaningful for more complicated structures.\\

The two types of transitions can notably be distinguished by the behavior of the fraction of surviving species $\omega_0$. As shown in Fig. \ref{fig:divergences}, this fraction is continuous at the instability for finite $\chi$ transitions, whereas infinite $\chi$ transitions are characterized by a sudden drop at the onset of instability. This qualitative difference can be related to the spectral properties of $A^\star$, discussed in \ref{sec:assembly_stability}. Indeed, as the reduced interaction matrix becomes unstable, the only possibility for the system to recover a \textit{new} stable fixed point is to kill enough species so as to recover a stable $A^\star$.  Whenever the spectral density is singular at the edge (that is, whenever $\chi\rightarrow\infty$), a non-zero fraction of the eigenvalues vanish at once. Hence the need to kill at least as many species as there are zero eigenvalues. This explains the discontinuous drop in $\omega_0$. On the contrary, if the spectrum of $A^\star$ vanishes at the edge (that is, if $\chi$ remains finite), stability can be achieved by killing one single species at a time, since the eigenvalues of the matrix become unstable one at a time. In that case, $\omega_0$ decreases continuously.

\section{Discussion}
\label{sec:disc}
Since their inception in theoretical physics, models with quenched disorder have proven a useful tool to understand complex systems in a wide range of fields. In ecology, they allow to bypass the need for finely-detailed biological information while providing non-trivial phenomenology. Yet, defining a random model requires deciding at which point in its construction randomness is added, and in which form. In turn, this implies strong assumptions that may both hinder the generality of the conclusions and limit the interpretability of the model.\\

In this work, we have provided a random framework that goes beyond previously studied cases and shown that a fundamental symmetry is enough to use the powerful tools of Dynamical Mean Field Theory and solve the dynamics in this more general scenario. Our symmetry, which requires the ensemble of interaction matrices to be invariant under orthogonal conjugation, is rooted in the idea that, in an unstructured ecosystem, the categorization of individuals into species is unrelated to the collective properties of the community. In other words, the community is defined by the macroscopic relations amongst its collective functions, rather than the association of specific species to specific functions. Due to this, the dynamics may not depend on individual interactions or species, but only in a few macroscopic quantities. In this \textit{collective regime}, the trajectories of species statistically decouple, and each of them can be seen as an independent realization of an effective stochastic process.\\

In more fundamental terms, orthogonally-invariant ensembles can also be seen as a solid and more general class of matrices to recover some of the results that have stood out as being shared across previous disordered models. In particular, the fact that SADs are truncated Gaussians can be directly related to the fact that scalar products are the only quantities that are preserved by orthogonal conjugation. Similarly, we have shown that the spectrum of the reduced interaction matrix and that of the full matrix are related by a transformation rule that generalizes previous results for Wigner matrices. This transformation is relevant for at least two reasons. First, it tells us that the surviving community, despite being dependent on the realization of the interactions, interacts in a way that is not so dissimilar from any typical sub-community of the same size, i.e. any subset of species sampled at random rather than obtained from the ecological dynamics. In other words, even though the extinction process might create subtle correlations and biases in the reduced interactions, these are not strong enough to yield a reduced spectrum that differs from the typical expectation. Even though our analysis focuses on the bulk, our numerical simulations and analytical calculations show no trace of outlier eigenvalues in the dynamics. Second, it tells us that smaller assembled communities (that is, communities with a smaller survival rate), tend to be more similar to the ones obtained from the Wigner model of interactions. Indeed, we have shown that the reduced spectrum interpolates between the full one and the semi-circle distribution, due to a faster decay of higher-order moments.\\

Within our framework, only two dynamical phases are possible. Either the system possesses a unique fixed point (UFP phase), which is reached regardless of initial conditions, or abundances diverge in finite time (DA phase). The analytical fixed point equations then generalize the findings of previous models, and make explicit the relationship of the spectral distribution with the properties of the equilibrium and the location of the instability. As for the diverging phase, even though it is ecologically meaningless, we argue that the addition of low-dimensional structure can replace it with more interesting regimes such as the Multiple Attractors phase described in previous works. In any case, the added structure must break the symmetry of the model. This paves the way for further work towards understanding how different ways of breaking the orthogonal symmetry are related to different types of ecological structure and in turn generate different types of dynamical regimes.\\

Furthermore, we believe that understanding how the orthogonally-invariant model transitions from the UFP phase to the DA phase can still be relevant to capture the local characteristics of the transitions that take place in more structured models. In particular, we show that the two types of diverging transitions present in our model (identified by the behavior of the response function $\chi$) preserve their distinctive features once a competitive bias in the interactions is added to regularize the divergences. This is because, as the mean abundance increases, the interaction bias decreases the growth rate of all the species at the same time. This curbs the divergence in the non-linear regime, but doesn't change the nature of the linear instability. Interestingly, we find that each kind of transition can thereby be mapped to a different behavior of the fraction of surviving species, which ties to the important ecological concept of coexistence. Infinite $\chi$ transitions lead to a sudden drop in the survival fractions, which could have a substantial impact in real ecological scenarios.\\

Crucially, all our calculations are dependent only on the spectrum of the interaction matrix. First, this establishes a powerful bridge with Random Matrix Theory, since spectra are typical outputs of RMT calculations. Additionally, by showing the equivalence between the instability of the fixed point solution and the instability of typical submatrices (which so far had only been addressed for Gaussian matrices \cite{Galla2023a}), we provide a solid justification for the use of RMT stability calculations. Second, and perhaps more importantly, our work paves the way towards a more thorough analysis of what characteristics of the spectrum are important to understand the dynamics. For example, we have shown that the behaviour of the spectrum at its edges plays a crucial role in determining the instabilities undergone by the system.\\

Finally, we wish to stress the potential benefits of grounding random models for community ecology on general considerations about the symmetries in species interactions. Here, orthogonal invariance has been used to translate the qualitative intuition that the dynamics of unstructured communities do not depend on the species’s identities. More generally, other symmetries may translate qualitative understandings of more complicated ecological scenarios, by differentiating between the characteristics that are relevant to the dynamics (and hence, must not be averaged out) and those that are not (and can be averaged out). The ensemble of matrices considered in Sec. \ref{sec:dyn_reg} can be seen as a very simple breaking of orthogonal symmetry into a slightly less stringent symmetry, whereby the total abundance of the community is singled out as a direction of the interaction matrix, and hence becomes relevant to the understanding of the dynamics. Likewise, further breaking the $\mathcal{O}_N$ symmetry would give rise to partially structured matrices generalizing those in \cite{poley2023}. More generally, starting from an altogether different symmetry group could translate a different ecological intuition and lead to a different pattern in the dynamics. To what extent the link made here between symmetries and patterns can be generalized to other symmetry groups is an open question that we believe is worth addressing in future work.
\begin{acknowledgments}
It is a pleasure to thank Guy Bunin, Tobias Galla, Jean-François Arnoldi, Matthieu Barbier and Silvia De Monte for insightful discussions and for useful comments during the preparation of this manuscript. The support of the Frontiers in Research and Education program and the EABIS Graduate Program in Life Sciences and Biodiversity is gratefully acknowledged.
\end{acknowledgments}
\bibliography{main}
\appendix
\section{Numerical simulations}
\label{sec:app_num}
All numerical calculations where performed in Python. Simulations where done with Scipy's \texttt{solve\_ivp} function with the \texttt{RK45} method.
Interaction matrices were sampled by fixing the spectral distribution and sampling an orthogonal matrix from the Haar measure. Initial conditions were
sampled uniformly between zero and 1. The solver was stopped after a fixed amount of time, generally $T=1000$. We used no immigration threshold.
Classification between the two possible phases is simplified by the fact that in the divergence phase the divergence happens in finite time. Numerically, this
causes Scipy's integrator to stop before $T$ is reached. Hence, we classify a trajectory as divergent if the solver stops before $T$ is reached.\\

Numerical solutions for the fixed point equations where obtained by using an iterative scheme. The equations where rewritten in vector form as $\Vec{X} = f(\vec{X})$ and
the iteration ran as $\Vec{X}_{n+1} = (1-r) \Vec{X}_n + r f(\Vec{X}_n)$ for a fixed number of iterations. We used $r=0.25$. The initial condition was taken to be $\Vec{X}_0 = \Vec{1}$.
Except very close to the divergence transition, we find that $200$ iterations were more than enough to converge to the solution, although the examples of the main text were solved with $20000$ iterations.
\section{The function $G_\rho$}
\label{sec:app_G}
The reader will find a more comprehensive review of the following notions in \cite{PottersBouchaud}.
For any random matrix $A$ with spectral distribution $\rho$, we define the Stieltjes transform of $\rho$ as
\begin{equation}
    \mathfrak{g}(x) = \int \frac{\rd \lambda \rho(\lambda)}{x-\lambda}.
    \label{eq:free_stieljes}
\end{equation}
$\mathfrak{g}$ is defined outside the support of the spectral distribution. For $x > \sup \mathrm{Supp}(\rho)$, $\mathfrak{g}$ is monotonically decreasing to zero, and is therefore invertible.
We can then define the $R$-transform as
\begin{equation}
    R_\rho(x) = \mathfrak{g}^{-1}(x) - \frac{1}{x}.
    \label{eq:def_R}
\end{equation}
The $R$-transform is defined over some interval $[0,\mathfrak{g}_\mathrm{max})$, where $\mathfrak{g}_\mathrm{max} = \lim_{x\rightarrow \sup \mathrm{Supp}(\rho)^+} \mathfrak{g}(x)$. One always has $R_\rho(0) = 0$.
We will make extensive use of the integral of the $R$-transform,
\begin{equation}
    G_\rho(x) = \int_0^x \rd y R_\rho(y).
\end{equation}
In particular, $G_\rho$ is the 'free probabilites analogue' of the cumulant generating function from free probabilities. Indeed, its series expansion is formally given by
\begin{equation}
    G_\rho(x) = \sum_n \frac{\kappa_n}{n!} x^n,
    \label{eq:G_series}
\end{equation}
where $\kappa_n$ are the so-called free cumulants of $\rho$. For our needs the existence of the cumulants is guaranteed by the fact that we'll only consider cases where $\rho$ is bounded. In particular,
one may take the previous series expansion and use it to extend the definition or $R$ and $G$ to the complex plane.\\

The examples used in the main text are:
\begin{enumerate}
    \item A Wigner distribution with variance $\sigma^2$.
    \item A Marcenko-Pastur distribution with ratio $\alpha$.
    \item A uniform distribution with edges $\pm\epsilon$.
    \item A binary distribution with half the eigenvalues at $-\lambda$ and the other half at $\lambda$.
\end{enumerate}
For these we have:
\renewcommand{\arraystretch}{2}
\begin{center}
    \begin{tabular*}{\linewidth}{@{\extracolsep{\fill}} ccc}
     \toprule
     \textbf{Type} & $G_\rho'$ & $G_\rho''$ \\ [1ex] 
     \toprule
     \textbf{Wigner} & $\sigma^2 x$ & $\sigma^2$ \\[1ex]
     \textbf{Hebbian} & $\frac{\alpha x}{1 + x}$ & $\frac{\alpha}{(1+x)^2}$ \\[1ex]
     \textbf{Uniform} & $\frac{\epsilon}{\tanh(\epsilon x)} - \frac{1}{x}$ & $\frac{1}{x^2} - \frac{\epsilon^2}{\sinh^2(\epsilon x)}$ \\[1ex]
     \textbf{Binary} & $\frac{\sqrt{1+4 \lambda^2 x^2} -1}{2x}$ & $\frac{\sqrt{1+4\lambda^2 x^2} -1}{2x^2 \sqrt{1+4\lambda^2 x^2}}$ \\[1ex]
     \hline
    \end{tabular*}
\end{center}
\section{Relationship between $G_\rho$ and the interaction coefficients}
\label{sec:app_coefs}
The joint distribution of a set of interaction coefficients $A_{i_1 j_1},\dots,A_{i_k j_k}$ can be characterized through their generating function
\begin{equation*}
    f(\lambda_1,\dots,\lambda_k) = \left\langle \exp\left(\lambda_1 A_{i_1 j_1} + \dots + \lambda_k A_{i_k j_k}\right)\right\rangle,
\end{equation*}
where the average is taken over realizations of the interaction matrix, \textit{i.e.} over realizations of the Haar measure. This average is an instance of the HCIZ integral from \eqref{eq:3_1_HCIZ} with $B = \sum_p \lambda_p \left(\ket{i_p}\bra{j_p} + \ket{j_p}\bra{i_p}\right)$. Assume that $k \ll S$, so that the HCIZ formula can be applied. We obtain
\begin{equation*}
    f(\lambda_1,\dots,\lambda_k) \sim \exp\left(\frac{S}{2} \mathrm{Tr} G_\rho\left(\frac{B}{S}\right)\right).
\end{equation*}
Assume for simplicity that the indices $i_p,j_p$ are disjoint between pairs with different $p$. Then the eigenvalues of $B$ are the $\pm \lambda_p$, and thus
\begin{equation*}
\begin{split}
    f(\lambda_1,\dots,\lambda_k) &\sim \exp\left(\frac{S}{2} \sum_p G_\rho \left(\frac{\lambda_p}{S}\right) \right)\\
    &\sim \exp\left(\frac{\lambda_p^2 \sigma^2}{2S} + \mathcal{O}(S^{-3})\right),
\end{split}
\end{equation*}
where $\sigma^2 = G_\rho''(0) = \langle\lambda^2\rangle$ is the variance of the spectrum. We recognize the generating function of a set of centered Gaussian variables with common variance $\sigma^2/S$ and two-point correlations of order at most $S^{-2}$. Thus, correlations between small sets of coefficients vanish.\\

Inter-dependencies between coefficients appear when an extensive number of them is considered at once. For example, let us study the sum of all interaction coefficients $\bra{1} A \ket{1}$. Its generating function is
\begin{equation*}
\begin{split}
    f(\lambda) = \left\langle \exp\Big(\lambda \bra{1}A \ket{1}\Big)\right\rangle &\sim \exp\left(\frac{S}{2} \mathrm{Tr} G_\rho\left(\frac{2\lambda\ket{1}\bra{1}}{S}\right)\right)\\
    &\sim \exp\left(\frac{S}{2} G_\rho(2\lambda)\right),
\end{split}
\end{equation*}
and therefore depends on the shape of $G_\rho$. In particular, the sum of the coefficients is not Gaussian, showing that the coefficients themselves are not jointly Gaussian. A Gaussian generating function is only recovered when the spectrum is semi-circular and $G_\rho(x) = \sigma^2 x^2 / 2$.
\newpage

\end{document}

\maketitle

\section{Properties of the interaction matrix}
\subsection{Scaling}
We aim to work in a regime where the number of extant species is extensive in the system size. Given the dynamical equation
\begin{equation}
    \frac{\rd x_i}{\rd t} = x_i \left(1 - u x_i + \sum_j A_{ij} x_j \right),
    \label{eq:scale_glv}
\end{equation}
this means that the term $A \ket{x}$ needs to be $\mathcal{O}(1)$ even in the presence of an extensive number of species
that all have abundances $\mathcal{O}(1)$. Given that the eigenvectors of $A$ are randomized by the action of $U$, for any
fixed $\ket{x}$ one has
\begin{equation}
    A \ket{x} = \sum_{\lambda \in \rho} \lambda \braket{v_\lambda | x}  \ket{v_\lambda},
    \label{eq:scale_dot}
\end{equation}
where the $v_\lambda$ are the random eigenvectors of $A$. One then has $\braket{v_\lambda | x} \sim \mathcal{O}(1)$ and $A \ket{x} \sim \mathcal{O}\left(\sqrt{\langle\lambda^2\rangle}\right)$.
The regime we seek is therefore attained when the spectrum does not scale with $S$.
\subsection{Spectrum}
The typical spectrum of a random matrix is composed of a bulk and outliers. The bulk accounts for most of the eigenvalues, and is responsible for the spectral
density of the matrix in the limit $S\rightarrow \infty$. The outliers are eigenvalues outside of the bulk, and do not contribute to the spectral density due to them being isolated.
In a regime of bounded spectrum like ours, bulk and outlier eigenvalues can be distinguished by their respective eigenvectors. Those corresponding to bulk eigenvalues are, broadly speking, random,
whereas those corresponding to outliers are non-typical.\\
Conjugating any matrix by a random orthogonal matrix doesn't change the spectrum, but effectively makes all eigenvectors random. Hence outlier eigenvectors cease to be non-typical and become indistinguishable
from bulk eigenvectors. Yet, because they are isolated, their corresponding eigenvalues cannot contribute to the spectral density. Hence, conjugating by a random orthogonal matrix effectively removes
outliers from the spectrum. This is expected for a model of unstructured communities, since outliers can be interpreted as structural modes of the system.

\section{Numerical simulations}
\subsection{Numerical simulations}
All numerical calculations where performed in Python. Simulations where done with Scipy's \texttt{solve\_ivp} function with the \texttt{RK45} method.
Interaction matrices were sampled by fixing the spectral distribution and sampling an orthogonal matrix from the Haar measure. Initial conditions were
sampled uniformly between zero and 1. The solver was stopped after a fixed amount of time, generally $T=1000$. We used no immigration threshold.
Classification between the two possible phases is simplified by the fact that in the divergence phase the divergence happens in finite time. Numerically, this
causes Scipy's integrator to stop before $T$ is reached. Hence, we classify a trajectory as divergent if the solver stops before $T$ is reached.\\
\subsection{Numerical solutions}
Numerical solutions for the fixed point equations where obtained by using an iterative scheme. The equations where rewritten in vector form as $\Vec{X} = f(\vec{X})$ and
the iteration ran as $\Vec{X}_{n+1} = (1-r) \Vec{X}_n + r f(\Vec{X}_n)$ for a fixed number of iterations. We used $r=0.25$. The initial condition was taken to be $\Vec{X}_0 = \Vec{1}$.
Except very close to the divergence transition, we find that $200$ iterations were more than enough to converge to the solution, although the examples of the main text were solved with $20000$ iterations.\\
\section{Obtaining the effective process with Dynamical Mean-Field Theory (DMFT)}
Here we describe the core calculations leading to the effective stochastic process. Readers not wanting to go through
them can jump to \ref{sec:summary_dmft} to obtain the definitions of the objects used in the main text.
\subsection{Free probabilities definitions}
\label{sec:free}
\paragraph{Free transforms}
We review very briefly the notions of free probabilities that will be useful in the coming sections. 
The reader will find a more comprehensive review of these notions in \cite{PottersBouchaud}.
For any random matrix $A$ with spectral distribution $\rho$, we define the Stieltjes transform
of $\rho$ as
\begin{equation}
    \mathfrak{g}(x) = \int \frac{\rd \lambda \rho(\lambda)}{x-\lambda}.
    \label{eq:free_stieljes}
\end{equation}
$\mathfrak{g}$ is defined outside the support of the spectral distribution. For $x > \sup \mathrm{Supp}(\rho)$, $\mathfrak{g}$ is monotonically decreasing to zero, and is therefore invertible.
We can then define the $R$-transform as
\begin{equation}
    R_\rho(x) = \mathfrak{g}^{-1}(x) - \frac{1}{x}.
    \label{eq:def_R}
\end{equation}
The $R$-transform is defined over some interval $[0,\mathfrak{g}_\mathrm{max})$, where $\mathfrak{g}_\mathrm{max} = \lim_{x\rightarrow \sup \mathrm{Supp}(\rho)^+} \mathfrak{g}(x)$. One always has $R_\rho(0) = 0$.
We will make extensive use of the integral of the $R$-transform,
\begin{equation}
    G_\rho(x) = \int_0^x \rd y R_\rho(y).
\end{equation}
In particular, $G_\rho$ is the 'free probabilites analogue' of the cumulant generating function from free probabilities. Indeed, its series expansion is formally given by
\begin{equation}
    G_\rho(x) = \sum_n \frac{\kappa_n}{n!} x^n,
    \label{eq:G_series}
\end{equation}
where $\kappa_n$ are the so-called free cumulants of $\rho$. For our needs the existence of the cumulants is guaranteed by the fact that we'll only consider cases where $\rho$ is bounded. In particular,
one may take the previous series expansion and use it to extend the definition or $R$ and $G$ to the complex plane.

\paragraph{HCIZ asymptotics}
In the following, we will be interested in integrals of quadratic exponentials over the orthogonal group $\mathcal{O}(S)$. These will take the form
\begin{equation}
    \mathcal{I}_S(\alpha,A,B) = \int \rd \mu(U) \exp\left(\frac{\alpha S}{2} \mathrm{Tr}\left[  U A U^T B \right]\right),
    \label{eq:def_HCIZ}
\end{equation}
where $\mu$ is the Haar measure over the orthogonal group and $A,B$ are symmetric matrices. These are the so-called Harish-Chandra-Itzykson-Zuber (HCIZ) integrals \cite{HCIZ}. Although no simple-enough general
expression exists for them, the following asymptotics are valid in the limit $S\rightarrow\infty$ with $\alpha$ finite\cite{Zuber,PMR}
\begin{equation}
    \mathcal{I}_S(\alpha,A,B) = \exp\left(\frac{S}{2} \mathrm{Tr} G_\rho(\alpha B)\right).
\end{equation}
These hold provided that the rank of $B$ is small compared to $S$ (we will always keep it finite) and that $\alpha$ is real and small enough so that the $R$-transform is defined at $\alpha$. Although
the formula is in principle not valid outside the latter condition, we find that the final formulas obtained for our particular problem are valid if one analytically extends the definition of $G_\rho$.
\subsection{DMFT}
\label{sec:dmft}
We now describe how to go from the deterministic Lotka-Volterra equations with random prescription for the interaction matrix
to the set of stochastic single-species processes with dynamical noise. Recalling that the trajectories of single species are random
due to the randomness in the interactions the idea is to express their probability distribution and simplify it to a point where it can
be recognized as the probability density associated to a stochastic process. We do so with the well-known framework of the Martin-Siggia-Rose path
integral formalism. The path integral associated to the Lotka-Volterra equations is simply
\begin{align}
    \zcal\left[\jvec\right] &= \left\langle \exp\left(i\sum_i \int \rd t j_i(t) x_i(t) \right)\right\rangle\\
    &= \int \rd A \mathbb{P}\left[A\right]\int \dcal \Vec{x}(t) \exp\left(i\sum_i \int \rd t j_i(t) x_i(t) \right) \prod_i \delta_f \left(\frac{\dot{x}_i}{x_i} -1 + u x_i - \sum_j A_{ij} x_j\right),
    \label{eq:dmft_path_integral}
\end{align}
where $\delta_f$ is a functional Dirac delta enforcing the Lotka-Volterra equations at each time step. We haven't explicitly represented the initial conditions as these don't play any role. They can be
fixed or random.
Seeing the functional delta as a product of deltas for each time step, we can use the Fourier representation
\begin{equation}
    \delta\left(x\right) = \int \frac{\rd \xh}{\sqrt{2\pi}} e^{-i \xh x},
    \label{eq:dmft_delta_fourier}
\end{equation}
so that the path integral becomes, up to a constant,
\begin{equation}
    \int \rd A \mathbb{P}\left[A\right]\int \dcal \Vec{x}(t)  \dcal \Vec{\hat{x}}(t)\exp\left(i\sum_i \int \rd t j_i(t) x_i(t) \right) \exp \left(-i\sum_i \int \rd t \xh_i(t) \left(\frac{\dot{x}_i}{x_i} -1 + u x_i - \sum_j A_{ij} x_j\right)\right),
    \label{eq:dmft_path_integral_fourier}
\end{equation}
The integral over the matrix coefficients can now be performed. Given the way our matrix ensemble is constructed,
we have,
\begin{align}
    \int \rd A \exp\left(i \sum_{ij} \int \rd t\xh_i(t) A_{ij} x_j(t)\right) &= \int \rd \mu(U) \exp\left(i \int \rd t \bra{\xh(t)} U A_0 U^T \ket{x(t)}\right)\\
    &= \int \rd \mu(U) \exp\left(\frac{i S}{2} \mathrm{Tr}\left[  U A_0 U^T \Sigma \right]\right),
    \label{eq:dmft_hciz}
\end{align}
where we have used the notation
\begin{equation}
    \Sigma = \frac{1}{S} \int \rd t \ket{x(t)}\bra{\xh(t)} + \ket{\xh(t)}\bra{x(t)} \approx \frac{\Delta t}{S} \sum_n \ket{x_i(n \Delta t)}\bra{\xh_i(n \Delta t)} + \ket{\xh_i(n \Delta t)}\bra{x_i(n \Delta t)}.
    \label{eq:dmft_sigma}
\end{equation}
In the r.h.s., the step $\Delta t$ is supposed to be small enough so as to provide an accurate approximation of the integral. Although we will
not bother with technicalities, one should consider that all time integrals have a finite upper bound (that will later be sent to $\infty)$, so
that, for finite $\Delta t$, the sum has a finite number of terms.\\
Hence for finite $\Delta t$, and if $S$ is sent to $\infty$, the matrix $\Sigma$ has low rank compared to $A_0$, and one can use the
HCIZ formula to calculate the integral over the orthogonal group. Note that this formula requires both $A_0$ and $\Sigma$ to be
symetric, this is the reason for which we have symmetrized the trace.\\
To deal with the following cumbersome calculations, we will require some notations. Let us denote
\begin{align}
    x^+_i(t) &= x_i(t) \\
    x^-_i(t) &= \xh_i(t),
    \label{eq:dmft_x_pm_def}
\end{align}
and introduce the tensor
\begin{equation}
    Q^{ab}(t,s) = \frac{\Delta t}{S} \sum_i x^{-a}_i(t) x^b_i(s). 
    \label{eq:dmft_Q_def}
\end{equation}
Note the sign flip for the $a$ index. Thus defined, $Q$ has two sets of indices, one for the $\{-,+\}$ and another for the time indices. It will therefore be convenient to
picture it as a $2\times 2$ block matrix, where each block corresponds to a combination of $,-+$, as in
\begin{equation}
    Q \equiv \begin{pmatrix}
        Q^{--} & Q^{-+} \\
        Q^{+-} & Q^{++}
    \end{pmatrix},
    \label{eq:dmft_Q_matrix_block}
\end{equation}
Each of the blocks is a matrix whose indices relate to time. Crucially, the following equality holds,
\begin{align}
    \mathrm{Tr}\left[Q^k\right] &= \frac{\Delta t^k}{S^k}\sum_{a_j,t_j,i_j : j<k} x_{i_1}^{-a_1}(t_1) x_{i_1}^{a_2}(t_2) x_{i_2}^{-a_2}(t_2) x_{i_2}^{a_3}(t_3)\dots x_{i_k}^{-a_k}(t_k) x_{i_k}^{a_1}(t_1)\\
     &= \frac{\Delta t^k}{S^k}\sum_{a_j,t_j,i_j : j<k} x_{i_1}^{a_2}(t_2) x_{i_2}^{-a_2}(t_2) x_{i_2}^{a_3}(t_3) x_{i_3}^{-a_3}(t_3) \dots x_{i_k}^{a_1}(t_1) x_{i_1}^{-a_1}(t_1) = \mathrm{Tr}\left[\Sigma^k\right],
    \label{eq:dmft_Q_equals_sigma}
\end{align}
which implies that for any analytical function $\mathrm{Tr}\left[f(Q)\right] = \mathrm{Tr}\left[f(\Sigma)\right]$. We are now ready to
apply the HCIZ integral,
\begin{equation}
    \int \rd \mu(U) \exp\left(\frac{i S}{2} \mathrm{Tr}\left[  U A_0 U^T \Sigma \right]\right) = \exp\left(\frac{S}{2} \mathrm{Tr} G_\rho(i\Sigma)\right) = \exp\left(\frac{S}{2} \mathrm{Tr} G_\rho(iQ)\right),
    \label{eq:dmft_hciz_Q}
\end{equation}
where $\rho$ is the eigenvalue distribution of $A_0$ and $G_\rho$ is the integral of the $R$-transform of that distribution, namely
\begin{equation}
    G_\rho(x) = \int_0^x \rd y \left(\mathfrak{g}^{-1}(y) - \frac{1}{y}\right),
    \label{eq:dmft_G_def}
\end{equation}
where $\mathfrak{g}$ is the Stieltjes transform of $\rho$. Note that at any moment one can recast $Q$ into a time-continuous
object by sending $\Delta t\rightarrow 0$. In that case, matrix multiplication becomes a time convolution, and analytical functions
applied on $Q$ become non-linear convolutional operators, e.g.
\begin{equation}
    f(Q)(t,s) = \sum_k \frac{f^{(k)}(0)}{k!}\int \rd u_1 \dots \rd u_{k-1} Q(t,u_1) \dots Q(u_{k-1},s),
    \label{eq:dmft_Q_convolution}
\end{equation}
where we have ommited the $\{-,+\}$ component for clarity. Given this correspondence, and keeping in mind that our
analysis is well-grounded because one can always go back to the discrete case to define things, we will switch between
the discrete and continuous notations at will.\\
With \eqref{eq:dmft_hciz_Q} in hand, the path integral no longer depends
on the individual interaction coefficients $A_{ij}$ but only on the eigenvalue distribution $\rho$ and the matrix $Q$. Hence we introduce
$Q$ as an integration variable using the Fourier expansion of functional Deltas as before,
\begin{equation}
    \int \dcal Q^{ab}(t,s) \dcal \hat{Q}^{ab}(t,s) \exp \left(i S \sum_{ab} \int \rd t \rd s \hat{Q}^{ab}(t,s)\left( Q^{ab}(t,s) - \frac{1}{S}\sum_i x_i^{-a}(t) x_i^b(s)\right)\right).
    \label{eq:dmft_Q_fourier}
\end{equation}
Plugging this and \eqref{eq:dmft_hciz} into the path integral and rearranging the terms finally yields the lengthy expression
\begin{align}
    \zcal\left[j\right] &= \int \dcal Q^{ab}(t,s) \dcal \hat{Q}^{ab}(t,s) \exp\left(\frac{S}{2} \mathrm{Tr} G_\rho\left(iQ\right) + iS \sum_{a,b}\int \rd t\rd s \hat{Q}^{ab}(t,s) Q^{a,b}(t,s)\right)\\
    &\times \exp \left[\sum_i\ln \left(\int \dcal x \dcal \xh \exp\left(-i \int \rd t \xh(t)\left(\frac{\dot{x}}{x} -1 +u x\right) - i \sum_{ab}\int \rd t\rd s x^{-a}(t) \hat{Q}^{ab}(t,s) x^b(s) + i \int \rd t j_i(t)x_i(t)\right)\right)\right].
\end{align}
Note that the second line contains the path integral of a single species, which now no longer depends on the index $i$ (other than through $j$) because all species
are identically coupled to the variables $\hat{Q}$. We may now compute the integrals over $Q$ and $\hat{Q}$ by using
th saddle-point method, given that all exponential factors are of order $S$. To do so, we need to set to zero the derivative of the
action w.r.t. $Q$ and $\hat{Q}$. Start with
\begin{align}
    \frac{\delta}{\delta Q^{ab}(t,s)} \mathrm{Tr} Q^k &= \sum_p \sum_{t_i, a_i : i\neq p} Q^{a_1a_2}(t_1,t_2)\dots Q^{a_{p-1}a}(t_{p-1},t)Q^{b a_{p+1}}(s,t_{p+1}) \dots Q^{a_k a_1}(t_k,t_1)\\
    &= k \left(Q^{k-1}\right)^{ba}(s,t).
    \label{eq:dmft_action_derivatives}
\end{align}
Note the order flip in the indices after the last equality. Extending this to the analytical function $G_\rho$,
\begin{equation}
    \frac{\delta}{\delta Q^{ab}(t,s)} \mathrm{Tr} G_\rho(iQ) = i G'_\rho(iQ)^{ba}(s,t).
    \label{eq:dmft_Q_derivative}
\end{equation}
Hence the derivatives of the full action are
\begin{align}
    \frac{\delta}{\delta Q^{ab}(t,s)} &= \frac{i}{2} G'_\rho(iQ)^{ba}(s,t) + i \hat{Q}^{ab}(t,s) = 0 \Rightarrow \hat{Q}^{ab}(t,s) = - \frac{1}{2}G'_{\rho}(iQ)^{ba}(s,t)\\
    \frac{\delta}{\delta \hat{Q}^{ab}(t,s)} &= i Q^{ab}(t,s) -i \langle x^{-a}(t) x^b(s) \rangle = 0 \Rightarrow Q^{a,b}(t,s) = \langle x^{-a}(t) x^b(s) \rangle.
    \label{eq:dmft_action_derivatives_final}
\end{align}
In writing \eqref{eq:dmft_action_derivatives_final}, we have used the notation
\begin{equation}
    \left\langle \mathcal{O}\left[x,\xh\right] \right\rangle = \frac{\int \dcal x \dcal \xh \mathcal{O}\left[x,\xh\right] \exp\left(-i \int \rd t \xh(t)\left(\frac{\dot{x}}{x} -1 +u x\right) - i \sum_{ab}\int \rd t\rd s x^{-a}(t) \hat{Q}^{ab}(t,s) x^b(s)\right) }
    {\int \dcal x \dcal \xh \exp\left(-i \int \rd t \xh(t)\left(\frac{\dot{x}}{x} -1 +u x\right) - i \sum_{ab}\int \rd t\rd s x^{-a}(t) \hat{Q}^{ab}(t,s) x^b(s)\right)}.
    \label{eq:dmft_path_avg}
\end{equation}
Plugging these in the path integral, we obtain
\begin{equation}
    \zcal[j] = \prod_i \int \dcal x(t)\dcal \xh(t) \exp\left( -i\int \rd t \xh(t)\left(\frac{\dot{x}}{x} -1 +u x\right) + \frac{i}{2} \sum_{ab}\int \rd t\rd s x^{-a}(t) G'_\rho(iQ)^{ab}(t,s) x^b(s) + i \int \rd t j_i(t) x_i(t)\right),
    \label{eq:dmft_path_integral_final}
\end{equation}
where $Q$ is to be obtained from \eqref{eq:dmft_action_derivatives_final}. Recalling the definition of the path integral, the
probability of a given trajectory $\Vec{x}(t)$ can now be readily recognized,
\begin{equation}
    \prod_i \int \dcal \xh_i(t) \exp\left( -i\int \rd t \xh_i(t)\left(\frac{\dot{x_i}}{x_i} -1 +u x_i\right) + \frac{i}{2} \sum_{ab}\int \rd t\rd s x_i^{-a}(t) G'_\rho(iQ)(t,s)^{ab} x_i^b(s)\right).
    \label{eq:dmft_effective_proba}
\end{equation}
First, this expression factors into a product over species. Therefore, the stochastic processes undergone by different species
are independent of each other. Second, the factors don't depend on the index $i$, hence all the stochastic processes are indentical
in law, i.e. species are statistically equivalent. Finally, the probability measure associated to any of these processes is the
same that was used in defining the average in \eqref{eq:dmft_path_avg}, hence the averages calculated this way are just averages with
respect to the effective stochastic process characterized by the probability of individual species in \eqref{eq:dmft_effective_proba}.\\
Let us now finish the simplification of the path integral. The first step is to recognize that
\begin{equation}
    Q^{+-}(t,s) = \langle \xh(t) \xh(s) \rangle = 0.
    \label{eq:dmft_hat_avg_is_zero}
\end{equation}
In other words, the matrix $Q$ is block-upper-triangular. This implies that $G'_\rho(iQ)^{+-} = 0$. The second is to
note the symmetry $Q^{--}(t,s) = Q^{++}(s,t)$, which implies that $G'_\rho(iQ)^{--}(t,s) = G'_\rho(iQ)^{++}(s,t)$. Hence the probability
measure of a single species becomes
\begin{equation}
    \int \dcal \hat{x}(t) \exp \left[-i\int \rd t \xh(t) \left(\frac{\dot{x}}{x} - 1 +u x - \int \rd s H(t,s) x(s)\right) - \frac{1}{2} \int \rd t\rd s \xh(t)C_\xi(t,s) \xh(s)\right].
    \label{eq:dmft_single_species_proba}
\end{equation}
where we have written for clarity $H(t,s) = G_\rho'(iQ)^{--}(t,s)$ and $C_\xi(t,s) = -i G'_\rho(iQ)^{-+}(t,s)$.
This is recognized as the Martin-Siggia-Rose path integral for the following stochastic process
\begin{equation}
    \dot{x}(t) = x(t) \left(1 - u x(t) + \int \rd s H(t,s) x(s) + \xi(t)\right),
    \label{eq:dmft_effective_process}
\end{equation}
where $\xi(t)$ is a Gaussian colored noise with correlator $\langle \xi(t) \xi(s) \rangle = C_\xi(t,s)$. Note that
the block-triangular nature of $Q$ allows to simplify a bit more the definitions of $H$,
\begin{equation}
    H(t,s) = G_\rho' (iQ^{--})(t,s).
    \label{eq:dmft_H_Cxi_def_simple}
\end{equation}
The definition of $C_\xi$ cannot be simplified that much, but we note that in any power of $Q$, th $-+$ block
contains one and only one factor $Q^{-+}$.\\
All in all, \eqref{eq:dmft_effective_process} is the outcome of DMFT: the original deterministic dynamics with random coefficients have been recasted into an effective one-dimensional stochastic
process whose parameters ($H$ and $C_\xi$) are to be determined from the outcome of the process. This circular relation
allows for a self-consistent determination of the parameters.\\
Unfortunately, the self-consistent relations are extremely cumbersome. Indeed, $H$ and $C_\xi$ are non-linear functionals of $Q$ in 
a space whose dimension is the number of time points of the system. Therefore, not only is an analytical solution out of reach, even a numerical solution would be computationally very costly.
\subsection{Summary of DMFT}
\label{sec:summary_dmft}
Effectively, each species undergoes a stochastic process independent of the others given by
\begin{equation}
    \dot{x}(t) = x(t) \left(1 - u x(t) + \int \rd s H(t,s) x(s) + \xi(t)\right),
    \label{eq:dmft_effective_process_summary}
\end{equation}
where $\xi(t)$ is a Gaussian colored noise with correlator $C_\xi(t,s)$. The parameters $H$ and $C_\xi$ are
self-consistently determined from the process itself  by means of the function $G_\rho$ defined as
\begin{equation}
    G_\rho(x) = \int_0^x \rd y \left(\mathfrak{g}^{-1}(y) - \frac{1}{y}\right),
    \label{eq:dmft_G_def_summary}
\end{equation}
where $\mathfrak{g}$ is the Stieltjes transform of the eigenvalue distribution $\rho$ of the matrix $A_0$.
This function is recasted into a convolutional operator acting on two-variable time dependent functions in
the natural way
\begin{equation}
    G_\rho(f)(t,s) = \sum_k \frac{G_\rho^{(k)}(0)}{k!} \int \rd u_1 \dots \rd u_{k-1} f(t,u_1) \dots f(u_{k-1},s).
    \label{eq:dmft_G_convolution_summary}
\end{equation}
Also define the overlaps
\begin{align}
    C(t,s) &= \langle x(t) x(s) \rangle \\
    K(t,s) &= i \langle x(t) \xh(s) \rangle.
    \label{eq:dfmt_Q_def_summary}
\end{align}
With these in hand one defines $H(t,s)$ as
\begin{equation}
    H(t,s) = G_\rho'(K)(t,s).
    \label{eq:dmft_H_def_summary}
\end{equation}
The definition of $C_\xi$ is more involved, as it cannot be written explicitly in terms of $G_\rho$,
\begin{equation}
    C_\xi(t,s) = \sum_k \frac{G_\rho^{(k)}(0)}{k!} \sum_{n<k} K^n \star C \star K_T^{k-n-1},
    \label{eq:dmft_Cxi_def_summary}
\end{equation}
where the powers are to be understood in the convolutional sense as before, as is the $\star$. Also, $\hat{q}_T (t,s) = \hat{q}(s,t)$
is the time-transpose of $\hat{q}$.

\section{Expressions of $G_\rho$ for the spectrum examples}
\label{sec:ex}
Only the derivatives $G_\rho'$ and $G_\rho''$ appear in the equations. We give the expression of $G_\rho$ as well as that of the derivatives. 
The spectral distributions are characterized by the following parameters:
\begin{enumerate}
    \item \textbf{Wigner:} Interaction variance, $\sigma^2$.
    \item \textbf{Hebbian:} Ratio of number of traits to number of species in the pool, $\alpha$.
    \item \textbf{Uniform:} Lower and upper edges of the spectrum, $\pm\epsilon$.
    \item \textbf{Binary:} Two Dirac masses accounting for half the eigenvalues each, located at $\pm\lambda$.
\end{enumerate}

\renewcommand{\arraystretch}{2}
\begin{center}
\scalebox{1.25}{
    \begin{tabular}{| c || c c c|}
    
     \hline
     \textbf{Type} & $G_\rho$ & $G_\rho'$ & $G_\rho''$ \\ [1ex] 
     \hline\hline
     \textbf{Wigner} & $\frac{1}{2}\sigma^2 x^2$ & $\sigma^2 x$ & $\sigma^2$ \\[1ex]
     \textbf{Hebbian} & $\alpha x - \alpha \ln(1+x)$ & $\frac{\alpha x}{1 + x}$ & $\frac{\alpha}{(1+x)^2}$ \\[1ex]
     \textbf{Uniform} & $\ln \left(\frac{\sinh(\epsilon x)}{\epsilon x}\right)$ & $\frac{\epsilon}{\tanh(\epsilon x)} - \frac{1}{x}$ & $\frac{1}{x^2} - \frac{\epsilon^2}{\sinh^2(\epsilon x)}$ \\[1ex]
     \textbf{Binary} & $\frac{1}{2}\left(\sqrt{1+4\lambda^2 x^2} - \ln\frac{1+ \sqrt{1 + 4 \lambda^2 x^2}}{2} - 1\right)$ & $\frac{\sqrt{1+4 \lambda^2 x^2} -1}{2x}$ & $\frac{\sqrt{1+4\lambda^2 x^2} -1}{2x^2 \sqrt{1+4\lambda^2 x^2}}$ \\[1ex]
     \hline
    \end{tabular}
}
\end{center}

\section{Obtaining the equations for the stationary case}
\label{sec:stat}
We now specialize the results of the previous section to the stationary case. This is enough to get rid of
the 'functional' aspect of the self-consistent equations, already allowing for an implementable numerical scheme
to solve the dynamics.\\
We define stationarity by requiring that one-point statistics are independent of time and two-point statistics
only depend on time differences. In particular, this implies that $H(t,s) = H(t-s)$ and $C_\xi(t,s) = C_\xi(t-s)$.
From the matrix point of view, stationarity is equivalent to matrices being circulant. From the functional point
of view, it implies that the problem can be greatly simplified by considering Fourier/Laplace transforms.\\
In this section, we use hats to denote Fourier transforms. This notation is different from the hatted conjugate variables used in \ref{sec:dmft}.
 Taking the Fourier transform in the definition of $H$ immediately yields the first relation
\begin{equation}
    \hat{H}(\omega) = G_\rho'(i\hat{Q}^{--}(\omega)),
    \label{eq:stat_H_fourier}
\end{equation}
where for each $\omega$, the Laplace transforms $\hat{Q}^{--}(\omega)$ and $\hat{H}(\omega)$ are now scalars. Hence this
relation has now been reduced to a set of independent scalar relations. In terms of the function $K$,
\begin{equation}
    \hat{H}(\omega) = G_\rho'(\hat{K}(\omega)).
    \label{eq:stat_H_fourier2}
\end{equation}
To generalize the second relation, we start by considering powers
\begin{equation}
    \left(Q^k\right)^{-+} = \sum_{n<k} \left(Q^{--}\right)^n Q^{-+} \left(Q^{++}\right)^{k-n-1} \Rightarrow  \left(Q^k\right)^{-+}(\omega) = Q^{-+}(\omega) \frac{Q^{++}(\omega)^k - Q^{--}(\omega)^k}{Q^{++}(\omega)-Q^{--}(\omega)}.
    \label{eq:stat_powers_fourier}
\end{equation}
Recalling the symmetry between $Q^{++}$ and $Q^{--}$, and plugging this expression in the analytical function $G_\rho$
yields
\begin{equation}
    \hat{C}_\xi(\omega) = \hat{Q}^{-+}(\omega) \frac{\mathrm{Im}\left[G_\rho'\left(i\hat{Q}^{--}(\omega)\right)\right]}{\mathrm{Im}\left[i\hat{Q}^{--}(\omega)\right]}.
    \label{eq:stat_Cxi_fourier}
\end{equation}
In terms of $K$ and $C$ this becomes
\begin{equation}
    \hat{C}_\xi(\omega) = \hat{C}(\omega) \frac{\mathrm{Im}\left[G_\rho'\left(\hat{K}(\omega)\right)\right]}{\mathrm{Im}\left[\hat{K}(\omega)\right]}.
    \label{eq:stat_Cxi_fourier2}
\end{equation}
\section{Obtaining the equilibrium equations}
\label{sec:fp}
We now suppose that the system goes to a fixed point. This implies that the correlator $C$ becomes a scalar, so that
$\hat{C}(\omega) \propto \delta(\omega)$. Hence the correlator $C_\xi$ also becomes a scalar, i.e. the dynamical
noise freezes, as expected. Based on this, we introduce the notations
\begin{align}
    \chi &= \lim_{t\rightarrow\infty} \int_0^t \rd s K(t-s) = \lim_{\omega\rightarrow 0} \hat{K}(\omega)\\
    \eta &= \lim_{t\rightarrow\infty} \int_0^t \rd s H(t-s) = \lim_{\omega\rightarrow 0} \hat{H}(\omega)\\
    \hat{C}(\omega) &= q \delta(\omega)\\
    \hat{C}_\xi(\omega) &= z \delta(\omega).
    \label{eq:ufp_eta_def}
\end{align}
The stationary equations then become
\begin{align}
    \eta &= G_\rho'\left(\chi\right)\\
    z &= q G_\rho''\left(\chi\right).
    \label{eq:ufp_eqs}
\end{align}
Going back to the dynamical equation for the single-species process, and setting the derivative to zero, we get
two possible equilibria
\begin{equation}
    x_\infty = 0 \text{ or } x_\infty = \frac{1 + \xi_\infty}{u-\eta}.
    \label{eq:ufp_x_eqs}
\end{equation}
We have denoted $\xi_\infty = \lim_{t\rightarrow\infty} \xi(t)$, which exists due to $C_\xi$ becoming
time-independent. The second equilibrium is feasible only if it is positive. As for the first one, one can prove that it is only stable
when the second is unfeasible. Hence, there is a single stable fixed point and the equilibrium abundances are given by
\begin{equation}
    x_\infty = \max\left(0, \frac{1 + \xi_\infty}{u-\eta}\right).
    \label{eq:ufp_sad}
\end{equation}
By definition, $\xi_\infty$ is now a Gaussian random variable with zero mean and variance $z$. \eqref{eq:ufp_sad}
allows us to get an expression for $q$ in terms of $\eta$ and $z$,
\begin{equation}
    q = \left\langle x_\infty^2 \right\rangle = \frac{z}{(u-\eta)^2} \int_{-\frac{1}{\sqrt{z}}}^\infty \frac{\re^{-y^2 /2}}{\sqrt{2\pi}}\left(y + \frac{1}{\sqrt{z}}\right)^2 = \frac{z}{(u-\eta)^2}\omega_2.
    \label{eq:ufp_q}
\end{equation}
To finish it off, we just need an expression for $\chi$. This can be obtained by linearizing the dynamical equation
and applying the definition of $K(t,s)$. All in all one gets
\begin{equation}
    \chi = \frac{1}{u-\eta} \int _{-\frac{1}{\sqrt{z}}}^\infty \frac{\re^{-y^2 /2}}{\sqrt{2\pi}} = \frac{\omega_0}{u-\eta}.
    \label{eq:ufp_chi}
\end{equation}
This concludes the derivation of the equilibrium relations of the main text.
\section{Interactions of the assembled community}
\label{sec:red-mat}
In the assembled community, some species remain and some go extinct. The fraction of surviving species $\omega_0$ is predicted by the DMFT equations. Hence the matrix restricted to
the surviving community is a $\omega_0 S \times \omega_0 S$ sub-matrix of the original interaction matrix. Because the equilibrium abundances are linked to the interactions, this matrix
is not just a randomly sampled sub-matrix. For instance, species experiencing less competition or more mutualism are more likely to survive.\\
Still, numerical simulations suggest that the spectral distribution of the reduced matrix is the same as that of a random sub-matrix. This had already been proven for the Gaussian case,
and was numerically observed in the Hebbian case. In the following, we give a heuristic explanation of why this is expected to hold in general. Consider the equilibrium equations,
\begin{align}
    \eta &= G_\rho'\left(\frac{\omega_0}{u-\eta}\right)\\
    z &= q G_\rho''\left(\frac{\omega_0}{u-\eta}\right)\\
    q &= \langle x^2 \rangle.
    \label{eq:assemb_ufp}
\end{align}
We may rewrite these equations as
\begin{align}
    \eta &= G_\nu'\left(\frac{1}{u-\eta}\right)\\
    z &= q^{+} G_\nu''\left(\frac{1}{u-\eta}\right)\\
    q^{+} &= \langle x^2 \rangle_{+} = q /\omega_0,
\end{align}
where we have defined $G_\nu'(x) = G_\rho'(\omega_0 x)$, i.e. $G_\nu(x) = \omega_0^{-1} G_\rho(\omega_0 x)$. Thus defined $G_\nu$ characterizes some spectral distribution $\nu$. With this
rewriting, the fixed point equations are now identical to those of a system with spectral distribution $\nu$ and surviving fraction $\omega_0 = 1$. Since the properties of the fixed point
are related to the extant community, this suggests that the matrix restricted to the extant community is characterized by the spectral distribution $\nu$ defined in this way.\\
Crucially, one shows that the equality $G_\nu(x) = \omega_0^{-1} G_\rho(\omega_0 x)$ defines the spectral distribution of any random sub-matrix of size $\omega_0 S$ of the original interaction matrix, i.e.
the spectral distribution of the reduced matrix is the same as that of a random sub-matrix. This latter claim is proven in the following.

\paragraph{$G_\nu(x) = \omega_0^{-1} G_\rho(\omega_0 x)$ characterizes the spectrum of any typical sub-matrix}
We prove this claim using the HCIZ integral. Let $P_0$ be a fixed projection matrix of size $S\omega_0 \times S$. For our particular case, $P_0$ is just a 'diagonal' matrix selecting the surviving species,
but what follows is true for any projection matrix. We are interested in the spectral distribution $\nu$ of the matrix $P_0 U A_0 U^T P_0^T$, where $U$ is a typical orthogonal matrix.\\
For fixed $U$ and $\alpha$ small enough, the HCIZ formula in $S\omega_0$ dimensions tells us that
\begin{equation}
    G_\nu (\alpha) = \frac{2}{S\omega_0} \ln\left[\int \rd \mu(V) \exp\left(\frac{S\omega_0 \alpha}{2} \bra{1} V P_0 U A_0 U^T P_0^T V^T\ket{1}\right)\right],
    \label{eq:assemb_HCIZ_log}
\end{equation}
where $\ket{1}$ is normalized. As it stands, $\nu$ depends on $U$. However, we expect that any typical $U$ will yield the same spectral distribution. Therefore, we may average over $U$ to get
\begin{equation}
    G_\nu (\alpha) = \frac{2}{S\omega_0} \int \rd \mu(U)\ln\left[\int \rd \mu(V) \exp\left(\frac{S\omega_0 \alpha}{2} \bra{1} V P_0 U A_0 U^T P_0^T V^T\ket{1}\right)\right].
    \label{eq:assemb_HCIZ_log_avg}
\end{equation}
Following standard techniques, we use the replica trick to deal with the logarithm and exchange the order of integration. The key is the replica identity
\begin{equation}
    \ln \left[\int \rd \mu(V) \dots\right] = \lim_{n\rightarrow 0} \frac{\left[\int \rd \mu(V)\dots\right]^n - 1}{n} = \lim_{n\rightarrow 0} \frac{\left[\int \rd \mu(V_1) \dots \rd \mu(V_n) \dots\right] -1}{n} ,
    \label{eq:assemb_replica}
\end{equation}
where in the last identity we suppose that the $n$-th power can be calculated for integer $n$ and analytically continued to real $n$ to calculate the limit. We therefore start by calculating the expression
for integer powers,
\begin{equation}
    \int \rd \mu(V_1) \dots \rd \mu(V_n) \int \rd \mu(U) \exp\left(\frac{S\omega_0 \alpha}{2} \sum_{a=1}^n \bra{1} V_a P_0 U A_0 U^T P_0^T V_a^T\ket{1}\right).
    \label{eq:assemb_hciz_replica}
\end{equation}
We may now do the $U$ integral with the help of the HCIZ formula in $S$ dimensions
\begin{align}
    \dots&=\int \rd \mu(V_1) \dots \rd \mu(V_n) \exp\left(\frac{S}{2} \mathrm{Tr} G_\rho\left[\alpha \omega_0\sum_{a=1}^n V_a^T\ket{1}\bra{1} V_a\right]\right)\\
    &= \int \rd \mu(V_1) \dots \rd \mu(V_n) \exp\left(\frac{S}{2} \mathrm{Tr} G_\rho\left[\alpha \omega_0 Q\right]\right).
    \label{eq:assemb_hciz_overlaps}
\end{align}
We have defined the $n\times n$ matrix of overlaps
\begin{equation}
    Q_{ab} = \bra{1} V^T_a V_b \ket{1},
\end{equation}
which we introduce in the integral using the Fourier representation of $\delta$ functions
\begin{equation}
    \int \rd Q \rd \hat{Q}\int \rd \mu(V_1) \dots \rd \mu(V_n) \exp\left(\frac{S}{2} \mathrm{Tr} G_\rho\left[\alpha \omega_0 Q\right] + i S \sum_{a,b} \hat{Q}_{ab}\Big[Q_{ab} - \bra{1} V^T_a V_b \ket{1}\Big]\right).
\end{equation}
We may now do the integrals over $Q$ and $\hat{Q}$ using the saddle method. Extremizing the exponent w.r.t. to $Q,\hat{Q}$ yields the saddle equations
\begin{align}
    i\hat{Q}_{ab} &= -\frac{1}{2} \frac{\partial}{\partial Q_{ab}} \mathrm{Tr} G_\rho(\alpha \omega Q) = -\frac{\alpha \omega_0}{2} G_\rho'(\alpha\omega_0Q)_{ab}\\
    Q_{ab} &= \frac{\int \rd \mu(V_a) \rd \mu(V_b) \bra{1} V_a^T V_b \ket{1}\exp\left(\frac{S\alpha \omega_0}{2} G_\rho'(\alpha\omega_0Q)_{ab} \bra{1} V_a^T V_b \ket{1}\right)}
    {\int \rd \mu(V_a) \rd \mu(V_b) \exp\left(\frac{S\alpha \omega_0}{2} G_\rho'(\alpha\omega_0Q)_{ab} \bra{1} V_a^T V_b \ket{1}\right)}.
    \label{eq:assemb_saddle_eqs}
\end{align}
For any fixed $n$, $Q$ and $\hat{Q}$ are functions of $\alpha$ and $\omega_0$ only through the product $\alpha \omega_0$. Therefore they are invariant under the transformation $\alpha \rightarrow \alpha \omega_0$
and $\omega_0 \rightarrow 1$. Plugging in the expression for $G_\nu$ yields
\begin{align}
    G_\nu (\alpha) = \frac{2}{S\omega_0} \lim_{n\rightarrow 0} \frac{ \exp\left(\frac{S}{2} \mathrm{Tr} G_\rho\left[\alpha \omega_0 Q\right] + iS \sum_{ab}Q_{ab} \hat{Q}_{ab}\right) \int \rd \mu(V_1) \dots \rd \mu(V_n) \re^{-i S \sum_{a,b} \hat{Q}_{ab}\bra{1} V^T_a V_b \ket{1}} - 1}{n}
    \label{eq:assemb_G_replica}
\end{align}
The expression inside the limit is only a function of $\alpha \omega_0$. Therefore
\begin{equation}
    G_\nu(\alpha,\omega_0) = \frac{1}{\omega_0} G_\nu(\alpha\omega_0,1) = \frac{1}{\omega_0} G_\rho(\alpha\omega_0),
    \label{eq:assemb_G_final}
\end{equation}
where we use the fact that for $\omega_0=1$ the sub-matrix is just the whole matrix and therefore $\nu=\rho$. This concludes the proof.
\section{Stability of the fixed point}
\label{sec:instab}
\subsection{General setting}
We now analyze the stability of the fixed point. Most of the work has essentially been done in \ref{sec:stat}. As in that section, hats are here used to indicate Fourier transforms.
The strategy is to linearize the dynamical equation around the fixed point and introduce a perturbation to the
fixed point abundances. The stability is then assessed by probing divergences in the Fourier transform of the equation.
This is equivalent to checking for the stability of the Jacobian of the system.\\
Let $x(t) = x_\infty + \delta x(t)$, where $\delta x(t)$ and linearize the equations,
\begin{equation}
    \frac{\rd}{\rd t} \delta x(t) = \delta x (t) \left(1 - \left(u-\eta\right) x_\infty  + \xi_\infty\right) + x_\infty \left(-\left(u-\eta\right)\delta x(t) + \delta \xi (t)\right).
    \label{eq:stab_linearized}
\end{equation}
Here, $\delta\xi(t)$ is a small perturbation to the DMFT noise that is induced by the perturbation to the abundances. In particular,
the self-consistent relations between $C_\xi$ and $Q$ can also be linearized,
\begin{equation}
    C_\xi(t,s) = C_\xi^\infty + \delta C_\xi(t,s) = G_\rho'\left(iQ + i \delta Q\right)^{-+}(t,s).
    \label{eq:stab_Cxi_perturbation}
\end{equation}
The perturbed regime is not stationary, therefore $\delta Q$ is not circulant. However, the unperturbed regime
is, therefore $Q$ is circulant. This is enough to take Fourier transforms. Denote $\ket{\omega}$ the Fourier vector
associated to the $\omega$ mode, i.e. $\ket{\omega}(t) = \exp(i\omega t)$. For any circulant matrix, these are eigenvectors
with corresponding eigenvalue equal to the $\omega$ mode of the Fourier transform of any of the matrices rows.
Then conjugating the previous relation
by $\bra{\omega}$ and $\ket{-\omega} = \bra{\omega}^\dagger$ yields
\begin{equation}
    \hat{C}_\xi^\infty \delta(\omega) + \langle \lvert\delta\xi (\omega)\rvert^2 \rangle  = \bra{\omega}G_\rho'(iQ+i\delta Q)^{-+} \ket{-\omega}.
    \label{eq:stab_Cxi_perturbation_fourier}
\end{equation}
In order to evaluate the r.h.s., start with the case where $G_\rho'$ is a power. In that case we get
\begin{equation}
    \left[\left(Q+\delta Q\right)^k\right]^{-+} = \left(Q^k\right)^{-+} + \sum_{n<k} \left(Q^{--}\right)^n \delta Q^{-+} \left(Q^{++}\right)^{k-n-1}.
    \label{eq:stab_power}
\end{equation}
Using the fact that $Q$ is a stationary matrix,
\begin{equation}
    \langle \lvert\delta\xi (\omega)\rvert^2\rangle = \frac{\mathrm{Im} \left[i\hat{Q}^{--}(\omega)\right]^k}{\mathrm{Im} \left[i\hat{Q}^{--}(\omega)\right]} \left\langle \lvert \delta x(\omega)\rvert^2\right\rangle.
    \label{eq:stab_power_fourier}
\end{equation}
This is readily generalized to the analytical function $G_\rho'$,
\begin{equation}
    \langle \lvert\delta\xi (\omega)\rvert^2\rangle = \frac{\mathrm{Im} G_\rho'\left[i\hat{Q}^{--}(\omega)\right]}{\mathrm{Im} \left[i\hat{Q}^{--}(\omega)\right]} \left\langle \lvert \delta x(\omega)\rvert^2\right\rangle
    =  \frac{\mathrm{Im} G_\rho'\left[\hat{K}(\omega)\right]}{\mathrm{Im} \left[\hat{K}(\omega)\right]} \left\langle \lvert \delta x(\omega)\rvert^2\right\rangle.
    \label{eq:stab_analytical_fourier}
\end{equation}
Now, go back to the linearized equation \eqref{eq:stab_linearized}. At the fixed point, some species are extinct, in which case
$x_\infty = 0$ and others are extant. For extinct species, the second term of the r.h.s. vanishes, and the first term
is negative. Therefore extinct species may not cause an instability. For extant species, the first term is zero, due to
the equilibrium abundance \eqref{eq:ufp_sad}. Hence the stability of the system may be examined by looking only at the second
term and only for extant species, i.e. we study the dynamics
\begin{equation}
    \frac{\rd}{\rd t} \delta x(t) = - x_\infty (u-\eta) \delta x(t) + x_\infty \delta \xi(t)
    \label{eq:stab_linearized_extant}
\end{equation}
with initial condition $\delta x(0)$. Given that $\xi$ is a Gaussian  process and that the equation  is linear, $\delta x$ is also a Gaussian process,
which we may characterize using its first two moments. As a linear transformation of it, the Laplace transform is also
a gaussian field, and we will find more convenient to characterize this one. Using hats to denote Fourier transforms,
\begin{equation}
    \left(\frac{i\omega}{x_\infty} + u-\eta\right) \delta \hat{x}(\omega) = \delta \hat{\xi}(\omega) + \delta x(0).
    \label{eq:stab_laplace}
\end{equation}
To characterize instability, we look for divergences in the mean and the variance of $\delta \hat{x}(\omega)$.Taking means in \eqref{eq:stab_laplace} yields
\begin{equation}
    \langle \delta \hat{x}(\omega)\rangle = \frac{x_\infty\delta x(0)}{i\omega + x_\infty\left(u-\eta\right)}.    
    \label{eq:stab_linearized_extant_mean}
\end{equation}
This cannot be divergent as long as $u-\eta>0$. Now consider the second moment and use the previously-derived self-consistent conditions,
\begin{align}
    \left\lvert  \frac{i\omega}{x_\infty} + u-\eta\right\rvert^2\langle \lvert\delta \hat{x}(\omega)\rvert^2\rangle &= \delta x(0)^2 + \langle |\delta \xi(\omega)|^2\rangle \\
    &= \delta x(0)^2 +\omega_0 \frac{\mathrm{Im} G_\rho'\left[\hat{K}(\omega)\right]}{\mathrm{Im} \left[\hat{K}(\omega)\right]} \left\langle \lvert \delta \hat{x}(\omega)\rvert^2\right\rangle.
    \label{eq:stab_linearized_extant_variance}
\end{align}
The factor $\omega_0$ (which acounts for extinct species) comes from the fact that the average acts on all species, but the linearized equation is valid
only for extant species. Solving and taking the limit $\omega\rightarrow 0$ yields
\begin{equation}
    \langle \lvert\delta \hat{x}(\omega)\rvert^2\rangle \sim_{\omega\rightarrow 0} \frac{\delta x(0)^2}{\left \lvert \frac{i\omega}{x_\infty} + u-\eta \right \rvert^2 - \omega_0 G_\rho''(\chi)}.
    \label{eq:stab_linearized_extant_variance_limit}
\end{equation}
This remains finite as long as the condition $\left(u-\eta\right)^2 - \omega_0 G_\rho''(\chi) = 0$ is not met.
Hence the stability condition is
\begin{equation}
    \left(u-\eta\right)^2 - \omega_0 G_\rho''(\chi) \geq 0 \textrm{ i.e. } \left(u-\eta\right)^2 - \omega_0 G_\rho''\left(\frac{\omega_0}{u-\eta}\right) > 0 .
    \label{eq:stab_condition}
\end{equation}
\subsection{Link with the spectrum of the assembled community}
\label{sec:mat-inst}
In this section, let $Y=u-\eta$. Then the onset of instability derived in \eqref{eq:stab_condition} may be rewritten
\begin{equation}
    \frac{\rd}{\rd Y} G_\nu'\left(\frac{1}{Y}\right) = -1,
\end{equation}
where we denote $\nu$ the spectrum of the surviving community, which we recall is defined by the relation
$G_\nu(x) = \omega_0^{-1} G_\nu(\omega_0 x)$. Using the relation between the R-transform and the Stieltjes transform this
is equivalent to
\begin{equation}
    \frac{\rd}{\rd Y} \mathfrak{g}^{-1}\left(\frac{1}{Y}\right) = 0 \Leftrightarrow \mathfrak{g}' \circ \mathfrak{g}^{-1} \left(\frac{1}{Y}\right)=\pm\infty.
\end{equation}
We stress out that here $\mathfrak{g}$ is the Stieltjes transform of the reduced matrix, not the full matrix. In particular, the last equality may only
hold at the edge of the reduced spectrum, i.e.
\begin{equation}
    \mathfrak{g}^{-1}\left(\frac{1}{Y}\right) = \lambda_\pm \Leftrightarrow u = \eta + \frac{1}{\mathfrak{g}\left(\lambda_\pm\right)},
    \label{eq:stab_assemb_chi}
\end{equation}
where we have denoted $\lambda_-,\lambda_+$ the edges of the spectrum. Combining this with the fixed point equation for $\eta$
we finally get
\begin{equation}
    u = \lambda_\pm.
    \label{eq:stab_assemb_condition}
\end{equation}
With our sign conventions for the interactions, this condition always happens at $u=\lambda_+$, at which point the spectrum of the
matrix $\mathbb{I} - A$ touches the imaginary axis. Hence the instability is signalled by the reduced interaction matrix.
\subsection{Finite $\chi$ and infinite $\chi$ transitions}
\label{sec:div-type}
Be it from the instability condition \eqref{eq:stab_condition} or from \eqref{eq:stab_assemb_chi}, one can see that the
marginality condition condition may happen either for $u-\eta \rightarrow 0$ or for $u-\eta > 0$. In the first case, the linear
response $\chi$ becomes infinite, whereas in the second case it stays finite.\\
Clearly, the difference between the two is marked by the behaviour of the Stieltjes transform a the edge of the reduced
spectrum. In particular, if $\mathfrak{g}(\lambda_+) = \infty$  then one has a $\chi\rightarrow\infty$ transition. Otherwise
a finite $\chi$ transition happens. More precisely, at the transition,
\begin{equation}
    \chi = \omega_0 \mathfrak{g}(\lambda_+).
\end{equation}\\
Given the definition of the Stieltjes transform, its behavior around $\lambda_+$ depends on the behavior of the spectral distribution
near the edge of the spectrum:
\begin{enumerate}
    \item If the spectral density at the edge goes to zero as $\rho(\lambda)=(\lambda_+-\lambda)^\delta$ for
    $\delta>0$ then the Stieljes transform stays finite at $\lambda_+$ and so does $\chi$. This is always the case for Wigner
    matrices since their submatrices are also Wigner and therefore have vanishing density at the edges of the spectrum.
    \item If the spectral density doesn't go to zero at the edge then the Stieltjes transform diverges and so does $\chi$. The
    simplest case scenario here is that of a specturm having a Dirac mass at the edge of the spectrum. This had already been
    observed numerically in the case of Hebbian interactions when the number of species in the surviving community is larger
    than the number of traits or ressources.
\end{enumerate}
To examine which transition is realized, one needs to understand the behaviour of $(u-\eta)^2 -\omega_0 G_\rho''(\chi)$ as $\chi\rightarrow\infty$. This may be rewritten as $\omega_0^2 / \chi^2 - \omega_0 G_\rho''(\chi)$.
The finite $\chi$ transition happens first if this expression is negative when $\chi$ is large but finite. This is equivalent to
\begin{equation}
    \omega_0 - \chi^2 G_\rho''(\chi) < 0 \textrm{ when } \chi \textrm{ is large but finite}.
    \label{eq:stab_chi_transition}
\end{equation}
Finally using the fact that the finite $\chi$ transition happens at $\omega_0=\omega_2=1/2$ we get the condition
\begin{equation}
    \frac{1}{2} < \lim_{x\rightarrow\infty} x^2 G_\rho''(x).
    \label{eq:stab_chi_transition2}
\end{equation}
If this condition is met, the finite $\chi$ transition happens before the other can take place. Otherwise, the $\chi\rightarrow\infty$ transition happens first.
This condition can in fact be given a purely algebraic meaning. Under the assumption that $\mathfrak{g}(\lambda_+) = +\infty$, the Stieltjes
transform is invertible over the whole positive real line, and it is a decreasing function. In particular, it can be inverted to yield a map
$\mathfrak{g}^{-1} :  (0,\infty) \rightarrow (\lambda_+,\infty)$ that is itself decreasing. Conversely, if the Stieltjes transform has a finite limit at
the edge, $\mathfrak{g}^{-1}$ is decreasing over $(0,\lambda_+)$. At $\mathfrak{\lambda_+}$, it reaches a critical point that signals the
branching between solutions to the inversion problem. Since we have the relation $G'_\nu(x) = \mathfrak{g}_\nu^{-1}(x) - 1/x$, it follows that the Stieltjes of
the reduced matrix remains finite at the edge of the spectrum iff there exists $x$ s.t. $G'_\rho(\omega_0 x) + 1/x$ is a critical point, before which the function is increasing
and after which it is decreasing. Therefore we need to have that for $x$ negative-enough $\omega_0 G''_\rho(\omega_0 x) - 1/x^2 = 0$, i.e.
$\lim_{x\rightarrow - \infty} x^2 G_\rho''(x) > \omega_0$. Using the fact that the finite transition always happens at $\omega_0=1/2$ we get the criterion
from the main text.\\

Note that this criterion also proves in passing that reducing the size of the matrix always has a smoothing effect, i.e. can
never cause a switch \textit{from} a finite transition \textit{to} an infinite transition. Indeed, if $\lim_{x\rightarrow - \infty} x^2 G_\rho''(x) > \omega_0$ holds
for some $\omega_0$ then it holds for any smaller $\omega_0$.
\subsection{All transitions are divergent transitions}
\label{sec:is-div}
Now that the instability threshold of the fixed point si established, we turn to the behavior of the system once
the threshold is crossed. Although the conclusion will always be the same, namely that the system enters a Divergent
Abundances phase in which it explodes in finite time, the analysis will be a bit different depending on the behavior
of $\chi$.\\
Using the equilibrium condition \eqref{eq:stab_condition} and the fixed point equations one has
\begin{equation}
    \frac{z}{q} \left(\omega_2 - \omega_0\right) = 0.
\end{equation}
This equality may be satisfied either by setting $\omega_2 = \omega_0$ or by sending $q/z \rightarrow \infty$:
\begin{enumerate}
    \item In the first case, one gets a finite $\chi$ transition. In particular, $\omega_2 = \omega_0$ implies that
    $\omega_0 = 1/2$ and $z\rightarrow \infty$, which in turn implies, using the fixed point equations, that $q\rightarrow\infty$.
    Hence this is indeed a transition characterized by a divergence of the abundances, accompanied by a divergence of the DMFT
    noise.
    \item In the second case, one gets an infinite $\chi$ transition. For this one,
    one no longer has $\omega_0=1/2$ and therefore $z$ remains finite. However, because $u-\eta \rightarrow 0$, one still
    has a divergence of $q$. Hence this is indeed a Divergent Abundances transition but one for which the intensity of
    the DMFT noise remains finite.
\end{enumerate}
\section{Extension to weakly sparse networks}
\label{sec:sparse}
We extend the previous calculations to interaction matrices with a weakly sparse topology. More precisely, the interaction matrix is built as the product
\begin{equation}
    A_{ij} = C_{ij} \Tilde{A}_{ij},
    \label{eq:sparse_ensemble}
\end{equation}
where the matrix $\Tilde{A}$ is sampled from an orthogonal ensemble as before and the matrix $C_{ij}$ is a \textit{connectivity matrix} with entries equal to
$0$ or $1$. For simplicity, we suppose that the entries are randomly sampled with probability $p$ of being equal to $1$, but some correlations could be included. DMFT is done in the same way as before
with the exception that we now need to average over both $C$ and $\Tilde{A}$. In other words, the relevant bit of the path integral now reads
\begin{equation}
    \int \rd\mathbb{P}(C) \int \rd \mu(U) \exp\left(i\mathrm{Tr}\left[  (U A U^T \otimes C)  \int \rd t \ket{x(t)}\bra{\xh(t)} \right]\right),
    \label{eq:sparse_hciz}
\end{equation}
where the operator $\otimes$ is the Hadamard product. It is convenient to carry out the integration over $C$ first. Indeed, we have
\begin{equation}
    \begin{split}
    \int \rd\mathbb{P}(C) \exp\left(i \mathrm{Tr}\left[  (U A U^T\otimes C) \int \rd t \ket{x(t)}\bra{\xh(t)} \right]\right) &= \int \rd\mathbb{P}(C) \prod_{ij} \exp\left(i (U A U^T)_{ij} C_{ij} \int \rd t x_j(t)\xh_i(t) \right) \\
    &= \prod_{ij} \left(1-p + p \exp\left(i (U A U^T)_{ij} \int \rd t x_j(t)\xh_i(t) \right)\right)\\
    &\approx \prod_{ij} \left(1 + i p (U A U^T)_{ij} \int \rd t x_j(t)\xh_i(t)\right)\\
    &\approx \exp\left(ip\sum_{ij} (U A U^T)_{ij} \int \rd t x_j(t)\xh_i(t)\right) \\
    &= \exp\left(\frac{ipS}{2} \mathrm{Tr}\left[U A U^T \Sigma\right]\right).
    \label{eq:sparse_Cint}
    \end{split}
\end{equation}
In the last line we have, similarly to \eqref{eq:dmft_hciz}, symmetrized the trace and introduced the matrix $\Sigma$ that was defined in \eqref{eq:dmft_sigma}. The HCIZ integral can now be performed,
and it is clear from \eqref{eq:sparse_Cint} that the only modification to all the results in the main text will be to replace $G_\rho(x) \rightarrow G_\rho(px)$.

\section{Extension to non-symmetric matrices}
\label{sec:asym}
\subsection{General setting}
We now extend the previous calculations to  non-symmetric matrices.
We consider the following ensemble
\begin{equation}
    A = \sqrt{\frac{1+\gamma}{2}}U A_+ U^T + \sqrt{\frac{1-\gamma}{2}}V A_- V^T,
    \label{eq:asym_ensemble}
\end{equation}
where $U,V$ are independent orthogonal matrices sampled from the Haar measure. $A_+$ is a fixed symmetric matrix
and $A_-$ is a fixed antisymmetric matrix. The parameter $\gamma$ controls the degree of symmetry of the ensemble.
In short, these are matrices of fixed symmetry coefficient $\gamma$ whose symmetric and antisymmetric parts are
independent. We characterize the symmetric part by its spectral density $\rho_+$ over the real axis and the antisymmetric
part by its spectral density $\rho_-$ over the imaginary axis. More precisely, we take $\rho_-$ to be (real) spectral
density of the Hermitian matrix $iA_-$.
\subsection{DMFT}
The DMFT analysis proceeds by performing the integrals over $U,V$ in a way similar to the symmetric analysis. By construction,
the integral over $U$ is exactly the same up to the prefactor. For the integral over $V$, the matrix $A_-$ is no longer symmetric, for this reason
we now antisymmetrize the abundance vector before performing the integral
\begin{align}
    \int \rd \mu(U) \exp\left(\frac{iS}{2}\sqrt{\frac{1+\gamma}{2}}\mathrm{Tr}\left[UA_+ U^T \int \rd t\left(\ket{x}\bra{\xh} + \ket{\xh}\bra{x}\right)\right] \right) &= \exp\left(\frac{S}{2} \mathrm{Tr} G_{\rho_+}\left(i\sqrt{\frac{1+\gamma}{2}} \Sigma_+\right)\right)\\
    \int \rd \mu(V) \exp\left(\frac{iS}{2}\sqrt{\frac{1-\gamma}{2}}\mathrm{Tr}\left[VA_- V^T \int \rd t\left(\ket{x}\bra{\xh} - \ket{\xh}\bra{x}\right)\right] \right) &= \exp\left(\frac{S}{2} \mathrm{Tr} G_{\rho_-}\left(\sqrt{\frac{1-\gamma}{2}} \Sigma_-\right)\right).\\
    \label{eq:asym_skew_integral}
\end{align}
Note that, in the second line, the factor $i$ is no longer present in the r.h.s. This is because $\rho_-$ is the spectral density
of $iA_-$. The matrix $\Sigma_+$ is the matrix $\Sigma$ of the symetric calculation, whereas $\Sigma_-$ is defined similarly with
a sign flip for the second term. Next one defines analogous $Q$ matrices, with $Q_+$ being identical to our previous $Q$ and $Q_-$
again taking the sign flip into account
\begin{equation}
    Q_-^{ab}(t,s) = \frac{\Delta t}{S} b \sum_i x_i^{-a}(t) x_i^b(s),
    \label{eq:asym_Q_minus}
\end{equation}
Note that $Q_-$ is a function of $Q_+$. Hence, to perform the saddle point as in the symmetric case, one needs to compute
the derivative of analytical functions of $Q_-$ with respect to the entries of $Q_+$. We find,
\begin{equation}
    \frac{\delta}{\delta Q_+ ^{ab}(t,s)} \mathrm{Tr} f(Q_-) = b f'(Q_-)_{ba}(s,t).
    \label{eq:asym_Q_diff}
\end{equation}
Therefore the saddle point equations read
\begin{equation}
    \hat{Q}_+^{ab}(t,s) = -\frac{1}{2} \left[\sqrt{\frac{1+\gamma}{2}} G_{\rho_+}' \left(i\sqrt{\frac{1+\gamma}{2}}Q_+\right)^{ba}(s,t) - ib \sqrt{\frac{1-\gamma}{2}} G_{\rho_-}'\left(\sqrt{\frac{1-\gamma}{2}}Q_-\right)^{ba}(s,t)\right].
    \label{eq:asym_saddle}
\end{equation}
Note that there is no variable $\hat{Q}_-$ since $Q_-$ has not been introduced as an independent quantity but
rather as a function of $Q_+$. As previously, $Q_+^{+-} = Q_-^{+-} = 0$ and therefore $\hat{Q}_+^{+-} = 0$. In particular, the structure of
the effective stochastic process is the same but the expressions of $H$ and $C_\xi$ are now more involved. For $H$,
\begin{align}
    H(t,s) &= \sqrt{\frac{1+\gamma}{2}} G_{\rho_+}' \left(i\sqrt{\frac{1+\gamma}{2}}Q_+^{--}\right)(t,s) + i\sqrt{\frac{1-\gamma}{2}} G_{\rho_-}'\left(\sqrt{\frac{1-\gamma}{2}}Q_-^{--}\right)(t,s)\\
    &= \sqrt{\frac{1+\gamma}{2}} G_{\rho_+}' \left(\sqrt{\frac{1+\gamma}{2}}K\right)(t,s) + i\sqrt{\frac{1-\gamma}{2}} G_{\rho_-}'\left(i\sqrt{\frac{1-\gamma}{2}}K\right)(t,s).
    \label{eq:asym_H_def}
\end{align}
For $C_\xi$, the expression is again harder to write in terms of $Q$ and $K$ so we will not write it down. Note that despite
appearances the antisymmetric term in \eqref{eq:asym_H_def} is real. Indeed $\rho_-$ must be symmetric, and therefore $G_{\rho_-}$ is
an odd function.
\subsection{Fixed points}
We now specialize the previous calculation to the fixed point phase, thereby extending the fixed point equations of the
symmetric case. The procedure is similar to that of the symmetric case, and exploits the circulant nature of the
matrices involved.
\begin{align}
    \eta &= \left[\sqrt{\frac{1+\gamma}{2}} G_{\rho_+}' \left(\sqrt{\frac{1+\gamma}{2}}\chi\right) + i\sqrt{\frac{1-\gamma}{2}} G_{\rho_-}'\left(i\sqrt{\frac{1-\gamma}{2}}\chi\right)\right]\\
    z &= q \left[\frac{1+\gamma}{2} G_{\rho_+}'' \left(\sqrt{\frac{1+\gamma}{2}}\chi\right) - i \chi^{-1} \sqrt{\frac{1-\gamma}{2}} G_{\rho_-}'\left(i \sqrt{\frac{1-\gamma}{2}}\chi\right)\right].
    \label{eq:asym_fp_eqs}
\end{align}

\printbibliography